\documentclass{aa}  
\usepackage{xcolor}
\usepackage{graphicx} 
\usepackage{physics}
\usepackage{amsmath}
\usepackage{amssymb}

\usepackage{natbib}
\bibpunct{(}{)}{;}{a}{}{,} % to follow the A&A style

\usepackage{threeparttable}

\usepackage[varg]{txfonts}
\usepackage{hyperref}
\usepackage{multirow}
\usepackage{color}
\definecolor{green}{rgb}{0.3,0.7,0.}
\definecolor{purple}{rgb}{0.77, 0.29, 0.55}

\hypersetup{
    colorlinks=true,       % false: boxed links; true: colored links
    linkcolor=blue,          % color of internal links (change box color with linkbordercolor)
    citecolor=blue,        % color of links to bibliography
    filecolor=blue,      % color of file links
    urlcolor=blue           % color of external links
}

\begin{document}

%\title{Dark matter annihilation in Pop III.1 supermassive stars}
\title{The evolution of protostars powered by dark matter annihilation. I. Fiducial model and first results}
\titlerunning{DM in SMS}
\author{Devesh Nandal\inst{1,2}, Konstantinos Topalakis\inst{3}, Jonathan C. Tan\inst{1,4}, Vasilisa Sergienko\inst{1}, Ana\"{\i}s Pauchet\inst{5}, Maya Petkova\inst{4}}
\authorrunning{Nandal et al.}

\institute{Department of Astronomy, University of Virginia, 530 McCormick Rd, Charlottesville, VA 22904, USA \and Center for Astrophysics, Harvard and Smithsonian, 60 Garden St, Cambridge, MA 02138, USA \and Department of Physics, University of Gothenburg, 412 96 Gothenburg, Sweden \and Dept. of Space, Earth \& Environment, Chalmers University of Technology, Chalmersgatan 4, 412 96 Gothenburg, Sweden \and I. Physikalisches Institut, Universität zu Köln, Zülpicher Str. 77, D-50937 Köln, Germany}

\date{}

\abstract{
The existence of billion-solar-mass quasars at redshifts $z \gtrsim 7$ poses a formidable challenge to theories of black hole formation, requiring pathways for the rapid growth of massive seeds. One such pathway may arise from primordial stars powered by dark matter (DM) self-annihilation rather than conventional hydrogen fusion, which could form massive black hole seeds in ordinary cosmological mini-haloes. Here we present a suite of stellar evolution models for DM-powered protostars, computed with the \textsc{GENEC} code. We systematically explored a wide parameter space, spanning ambient WIMP densities of $\rho_\chi \sim 10^{12}\mbox{--}10^{16}\,\mathrm{GeV\,cm^{-3}}$ and gas accretion rates of $10^{-3}\mbox{--}10^{-1}\,M_\odot\,\mathrm{yr^{-1}}$, to quantify the effects of DM annihilation.
A central finding is that for a protostar to grow to supermassive scales ($\gtrsim 10^5 \, M_{\odot}$), the ambient DM density in the immediate vicinity of the star must exceed a critical threshold of $\rho_{\chi} \gtrsim 5 \times 10^{14} \, \text{GeV cm}^{-3}$. The energy injected by WIMP annihilation inflates the protostar, lowering its surface temperature, which suppresses the ionising feedback that would otherwise halt accretion and significantly delays the onset of hydrogen fusion. This heating also governs the star's final fate; In dense halos ($\rho_\chi \gtrsim 10^{15}\,\mathrm{GeV\,cm^{-3}}$), stars remain stable against general relativistic instability beyond $10^6 \, M_{\odot}$, whereas at lower densities ($\rho_\chi \lesssim 10^{13}\,\mathrm{GeV\,cm^{-3}}$), they collapse at masses of $\sim 5 \times 10^5 \, M_{\odot}$. Once the DM fuel is exhausted and core burning commences, the protostar contracts and its ionising photon output can reach very high levels $\sim 10^{53}\:{\rm s}^{-1}$. These distinct evolutionary phases offer clear observational signatures for the \textsc{JWST}, providing a robust, physically grounded pathway for forming heavy black hole seeds in the early Universe.

}


\maketitle

\section{Introduction}

The rapid emergence of billion-solar-mass quasars within a few hundred million years after the Big Bang demands a robust explanation for the efficient formation and growth of supermassive black hole (SMBH) seeds in the early Universe \citep{Fan2003,Mortlock2011,Wang2021,Yang2021,Bogdan2024}. Any successful seeding scenario must also account for the apparent dearth of intermediate-mass black holes (IMBHs) in the local Universe \citep[e.g.][]{2025MNRAS.541..429M}. Various SMBH formation scenarios have been proposed, broadly categorised into `light seed' and `heavy seed' models \citep[e.g.][]{Volonteri2010,Greene2020}. Light seed models involve stellar remnants with masses $\sim10-100\:M_\odot$, which then grow rapidly via sustained Eddington or even super-Eddington accretion. On the other hand, heavy seed models invoke monolithic collapse of pristine gas clouds into supermassive `stars', i.e., with masses $\sim10^4-10^5\:M_\odot$, which then evolve to produce SMBHs. The most popular model of monolithic collapse to form heavy seeds is known as`direct collapse', and involves suppression of fragmentation in uv-irradiated or strongly turbulent metal-free relatively massive ($\sim 10^7\:M_\odot$), atomically cooled dark matter (DM) halos \citep[e.g.][]{Begelman2006,Omukai2008,Latif2013,Chon2016,Wise2019,2022Natur.607...48L,2025OJAp....8E..88O}. It is hypothesised that supermassive star formation is enabled by very high accretion rates to a central protostar, leading eventually to the seeding of a direct collapse black hole (DCBH).

When conditions for DCBH seeding do arise, collapse is hypothesised to lead to the formation of metal-free, supermassive stars (SMSs) under sustained, very high accretion rates \citep{Hosokawa_2012,Whalen2020,Latif_2014b}. One-dimensional stellar-evolution calculations show that rapidly accreting primordial protostars can remain in a bloated `supergiant protostar' state with relatively cool photospheres, which weakens ionising feedback and can permit growth to $\sim10^{4}$--$10^{5}\,M_\odot$ before the onset of the general relativistic instability and collapse to a massive black hole seed
\citep{Hosokawa_2013,Woods2017,Lionel2018,Herrington2023,Nandal2024c, Nandal2025d}.
Such SMS-based heavy-seed channels have been explored across a range of halo environments and accretion histories \citep{Latif_2014c,Wise2019,Regan_2020}. 

In conventional Population III star formation, H$_2$ cooling leads to baryon collapse at z $\approx$ 20 in small, pristine DM halos at rates of 10$^{-4}$ - 10$^{-3}$ $M_{\odot}$ yr$^{-1}$ when the halo reaches masses of ~ 10$^5$ $M_{\odot}$ \citep[e.g.][]{abn02,get11,clark11,susa13,hir15,latif22a}.  These infall rates produce a protostar that contracts and triggers p-p burning and then triple alpha and CNO burning, which stabilise it against further collapse.  However, if DM and baryons instead collapse adiabatically so that their respective density peaks always coincide, DM self-annihilation may provide enough energy to halt the contraction of the protostar before the onset of the p-p chain, creating a star powered by DM burning rather than nuclear fusion.  Such DM stars can grow to very large masses even at the modest accretion rates in cosmological minihalos because (1) their photospheres remain cool and red and do not produce the ionising UV flux required to terminate accretion onto the star and (2) DM self-annihilation replenished by cosmological flows never leads to the collapse of the core.  In this manner it may be possible for even normal minihalos to produce massive BH seeds and 'little red dots' \citep[LRDs;][]{lrd1,lrd2,lrd3,lrd4}, overmassive black hole galaxies \citep[e.g.][]{latif26a}, or even quasars \citep[e.g.][]{latif22b}.

Seeding via Population III.1 (Pop III.1) stars \citep{Banik2019,Singh2023,2025MNRAS.536..851C,Tan2024} is another promising model for generating the cosmic population of SMBHs. Pop III.1 stars are defined as being metal-free and forming from the first dark matter minihalo ($\sim10^6\:M_\odot$) but later collapsing in a given local region of the Universe such that star is unaffected by feedback, especially ionisation, from any neighbouring astrophysical source \citep{McKeeTan2008}. Such halos have traditionally been considered to form stars with masses of $\sim100-10^3\:M_\odot$, which form at best only 'light-seed' mass black holes, with this mass set by the point at which the stars contract to near the zero-age main-sequence (ZAMS) structure, resulting in strong ionising feedback \citep[e.g.][]{abn02,Bromm2002,Tan2004,McKeeTan2008,2010AIPC.1294...34T,2011Sci...334.1250H,Susa2014,Hirano2014}. Pop III.2 stars are defined as being those forming in metal-free minihalos that have been irradiated, leading to enhanced free electron abundances, which then catalyze increased abundances of $\rm H_2$ and HD coolants and lead to fragmentation in these minihalos to even lower-mass stars, i.e., with $\sim 10\:M_\odot$ \citep[e.g.][]{2006MNRAS.373..128G}. However, it has been proposed that WIMP annihilation heating, boosted to significant levels by adiabatic contraction of the dark matter density in Pop III.1 minihalos, could have a major impact on the formation of stars in these halos \citep[][]{Spolyar2008,2009ApJ...692..574N}. Energy injection from WIMP self-annihilation can act as a source of fuel to support the protostar in a configuration that is relatively large and with a relatively cool photospheric temperature \citep{Freese2010,Ilie2012,RindlerDaller2015,Ilie2021}, which could thus prevent strong ionising feedback and allow efficient growth of the star from baryonic content of the minihalo \citep{Tan2024}. In this context, the terms Pop III.1 and `dark star'  describe complementary aspects of the same physical scenario where Pop III.1 identifies the pristine, externally unirradiated minihalo environment, while dark star identifies the phase in which WIMP annihilation, rather than nuclear burning, provides the dominant support.

Previous models of these stars, referred to as `dark stars' \citep{Freese2010,RindlerDaller2015,Ilie2021}, have established that WIMP annihilation heating can support very massive, luminous (near Eddington luminosity) primordial objects in extended, relatively cool configurations. Many of these calculations focused on the continuous-support limit, in which the WIMP reservoir is efficiently replenished by the surrounding minihalo and dark matter heating remains important throughout the main growth phase. In this study, we follow the time dependent stellar response to WIMP capture and annihilation in \textsc{GENEC}, include the resulting ionising photon production and radiative feedback, and examine whether the objects encounter the general relativistic radial instability (GRRI) during their subsequent evolution \citep{Chandrasekhar_1964b,Baumgarte1999,Woods2017}.

Beyond black-hole seeding, SMSs formed either via the DCBH or DM annihilation seeding mechanism, may also leave observable chemical signatures. For example, it has been argued that very massive primordial stars can reproduce the unusually high N/O ratios reported in some high-$z$ galaxies observed by JWST \citep{Nandal2024,Nandal2025}.

In this work, we present a systematic numerical approach 
%to address these gaps 
by integrating Gould's robust DM capture formalism \citep{Gould1987} into the SMS branch of the \textsc{GENEC} stellar‑evolution code \citep{Eggenberger2008, Nandal2024d, Nandal2024c}.  The details of the capture implementation and its coupling to the stellar structure equations are given in Section~\ref{sec:methods_dm}, while the numerical set‑up and model grid are summarised in Section~\ref{sec:setup}.  By rigorously coupling DM annihilation heating with the baryonic microphysics already present in \textsc{GENEC}, we explore a broad parameter space in ambient WIMP density ($\rho_\chi = 10^{12}$–$10^{16}\,\mathrm{GeV\,cm^{-3}}$) and accretion rate ($\dot{M}_{*}=10^{-3}$–$10^{-1}\,M_\odot\,\mathrm{yr^{-1}}$).  The resulting evolution of stellar structure, luminosity budgets, and pre‑main‑sequence tracks are presented in Section~\ref{Sec:Transport}, followed by an analysis of ionising‑photon production and radiative feedback in Section~\ref{Sec:Ionising}.  Section~\ref{Sec:GR} quantifies the impact of DM heating on general‑relativistic (GR) stability, identifying density–dependent thresholds that either trigger or suppress collapse.  A comparison with previous theoretical and observational studies is provided in Section~\ref{Sec:CompPrev}, and broader implications, including potential JWST observables are discussed in Section~\ref{Sec:Discussion}.  Collectively, these results offer comprehensive insight into how DM environments shape the evolution, stability, and observational signatures of Pop~III.1 DM-powered stars, yielding concrete, testable predictions for forthcoming high‑redshift surveys. 

\section{Methods} \label{Sec:Methods}

%---------------------------------------------------------------
%  METHODS SECTION : POP III.1 DARK–MATTER CAPTURE MODULE
%---------------------------------------------------------------

\label{sec:methods_dm}

Protostars born inside DM minihalos are expected to be immersed in high DM densities. If the DM consists of WIMPs, 
%elastic 
scattering on nuclei leads first to gravitational capture, then to rapid thermalisation, and finally to self-annihilation of the captured particles in the stellar core.  
The associated heat source may strongly affect the structure and evolution of the stars. In this section we outline the physical model implemented in the \textsc{GENEC} stellar evolution code.  The treatment follows the original capture formalism of \citet{Gould1987} and the Pop III extensions of \citet{Taoso2008, Pauchet2026}, but capture is generalised here to time-dependent capture, annihilation, and (optionally) self-capture and evaporation.

%-----------------------------------------------------------------
%\subsection{Population evolution}
\subsection{Evolution of dark matter inventory}
\label{sec:pop_evol}

We began by tracking the total number of bound WIMPs in the star, \(N_\chi(t)\), which evolves according to
\begin{equation}
\frac{dN_\chi}{dt}
\;=\;
C_c                       % nuclear capture
\;+\;
C_{\mathrm{self}}\,N_\chi % self-capture
\;-\;
A\,N_\chi^{2}             % annihilation
\;-\;
E\,N_\chi,                % evaporation
\label{eq:dNdt}
\end{equation}
where \(C_c\) is the nuclear capture rate; \(C_{\mathrm{self}}\) is the self-capture rate; and \(A=\langle\sigma_a v\rangle/V_{\mathrm{eff}}\) is the annihilation coefficient with $\langle\sigma_a v\rangle$ as the thermally averaged annihilation cross section, $V_{\rm eff} = \sqrt{2}\,\pi^{-3/2}\,r_\chi^{3}$ as the effective volume, and \(E\) as the evaporation rate.  
Throughout this work we neglect self-capture and evaporation (\(C_{\mathrm{self}}=E=0\)) because, for the weak-scale cross-sections considered, self-capture is insignificant \citep[a more detailed work on DM self-capture has been conducted by][]{Zentner2009} and evaporation is exponentially suppressed for WIMP masses \(m_\chi\gtrsim5\)–10 GeV \citep{Gould1987}. Consequently the competition between \(C_c\) and \(A\,N_\chi^{2}\) fully determines \(N_\chi(t)\). This equation can be solved analytically, and the general solution is 
\begin{equation}
    N_x(t) = \sqrt{\frac{C_c}{A}} \tanh\left( \sqrt{C_c A} \, t \right).
\end{equation}
To close Eq.~\eqref{eq:dNdt} we now describe the calculation of \(C_c\), the spatial profile that defines \(V_{\mathrm{eff}}\), and the characteristic timescale on which equilibrium is reached.

%-----------------------------------------------------------------
\subsection{Nuclear capture in a mass shell}
\label{sec:capture}

The nuclear capture rate derives from single-scatter kinematics integrated over the WIMP halo velocity distribution and the local stellar structure.  For a shell of mass \(dm\) at radius \(r\), it is
\begin{equation}
\frac{dC}{dm} =
\sqrt{\frac{6}{\pi}}\,
\frac{\sigma_{\rm eff}\,\rho_\chi}{m_\chi}\,
\frac{v_{\rm esc}^{2}}{v_\chi}\,
\frac{\mathcal{P}}{2\sqrt{3/2}\,A^{2}},
\label{eq:dcdm}
\end{equation}
where \(\rho_\chi\) is the ambient DM density, \(v_\chi\) the one-dimensional halo velocity dispersion, \(v_{\rm esc}(r)\) the local escape speed, and \(\sigma_{\rm eff}=\sigma_{\mathrm{SI}}^{A}+\sigma_{\mathrm{SD}}^{A}\) the effective WIMP–nucleus cross-section (spin-independent plus spin-dependent).  The factor \(\mathcal{P}\) encloses the angular integration over Maxwellian halo velocities,
\begin{equation}
\begin{split}
\mathcal{P} &=
 (A_+A_- - 0.5)\,(1.62487-\chi_\pm) \\[4pt]
            &\quad{}+ 0.5\,A_+\,e^{-A_-^{2}}
                     - 0.5\,A_-\,e^{-A_+^{2}}
                     - \sqrt{\tfrac{3}{2}}\,e^{-3/2},
\end{split}
\label{eq:parent}
\end{equation}
with
\begin{align}
A^{2} &=
\frac{3\,v_{\rm esc}^{2}\,\mu}
     {2\,v_\chi^{2}\,\mu_{\rm red}^{2}},                    \label{eq:A2}\\[2pt]
A_\pm &=
\sqrt{A^{2}}\,
\pm
\sqrt{\tfrac{3}{2}}\,(v_*/v_\chi),                          \label{eq:Apm}\\[2pt]
\chi_\pm &=
0.88623\,[\,\erf(A_+)-\erf(A_-)\,].                         \label{eq:chipm}
\end{align}
Here \(v_*\) is the stellar bulk speed through the halo (taken to be negligible for minihalo stars), \(\mu=m_\chi/m_N\) is the WIMP–to–nucleus mass ratio, and \(\mu_{\rm red}=m_\chi m_N/(m_\chi+m_N)\) is the corresponding reduced mass.  
Equation \eqref{eq:A2} shows that \(A^{2}\) captures the kinematic suppression of capture when \(m_\chi\) greatly exceeds the nuclear mass \(m_N\) or when \(v_\chi\gg v_{\rm esc}\).  
We evaluate Eq.~\eqref{eq:dcdm} for every isotope and mesh point; trapezoidal summation yields the global capture rate \(C_c(t)\) that feeds back into Eq.~\eqref{eq:dNdt}.

%-----------------------------------------------------------------
\subsection{Thermalisation and spatial distribution}
\label{sec:profile}

After capture a WIMP scatters repeatedly, losing energy until it attains the local Maxwellian with a characteristic thermalisation time \(t_{\rm th}\sim10^{1}\)--\(10^{3}\,\mathrm{yr}\), much shorter than any nuclear or transport timescale.  
The resulting steady state is well described by an isothermal Gaussian profile:
\begin{equation}
n_\chi(r) =
n_{\chi0}\,
\exp\!\bigl(-r^{2}/r_\chi^{2}\bigr),
\qquad
r_\chi^{2} =
\frac{3\,k_{\rm B}T_c}{2\pi G\rho_c m_\chi},
\label{eq:rchi}
\end{equation}
where \(T_c\) and \(\rho_c\) are the instantaneous core temperature and density.  
The small scale radius \(r_\chi\) (typically \(10^{9}\)--\(10^{10}\,\mathrm{cm}\)) implies that annihilation heating is confined to the very centre of the star.

%-----------------------------------------------------------------
\subsection{Characteristic timescale}
\label{sec:tauchi}

Substituting \(C_c\) and the volume integral of \(n_\chi^2\) into Eq.~\eqref{eq:dNdt} defines the capture–annihilation equilibrium time;
\begin{equation}
\tau_\chi =
\bigl(C_cA\bigr)^{-1/2}
=
\sqrt{\frac{V_{\rm eff}}{C_c\,\langle\sigma_a v\rangle}}.
\label{eq:tauchi}
\end{equation}
For the fiducial parameters adopted below, \(\tau_\chi\approx10^{2}\,\mathrm{yr}\); hence the WIMP reservoir reaches its equilibrium value \(N_{\chi,\infty}=\sqrt{C_c/A}\) well before nuclear burning commences.

%-----------------------------------------------------------------
\subsection{Annihilation luminosity}
\label{sec:lumin}

In steady state, the volumetric annihilation rate is \(\Gamma_{\rm ann}=A\,N_\chi^{2}/2\).  
We assume one third of the annihilation energy is carried away by neutrinos that escape the core \citep{Scott2009}. Thus the net luminosity deposited in the star is
\begin{equation}
L_\chi = \frac{2}{3}\,m_\chi\,C_c,
\label{eq:Lchi}
\end{equation}
and the corresponding local heating rate is \(\epsilon_\chi(r)=L_\chi\,n_\chi^{2}(r)/[\rho(r)\,N_\chi^{2}]\).  
This term is added to the nuclear energy generation rate in the stellar structure equations. Having outlined in the preceding subsections the adopted microphysics and the dark–matter capture formalism, we now turn to the numerical engine that brings these ingredients together within \texttt{GENEC}.
% ------------------------------------------------------------------
\subsection{Numerical framework of \texttt{GENEC} and its coupling to dark matter capture}

\texttt{GENEC} advances a stellar model by simultaneously solving the four ordinary differential equations of one–dimensional structure with a classical Henyey relaxation scheme.  At the beginning of each timestep provisional profiles for pressure, temperature, luminosity, and radius are linearised; the resulting banded Jacobian is inverted so that the central and surface boundary conditions are met in a single global sweep.  Convergence is achieved when the relative corrections to all four variables fall below the Henyey tolerance, guaranteeing that hydrostatic equilibrium, energy conservation, and radiative/convective transport are satisfied to machine precision \citep{Nandal2024b}.

The freshly converged density $\rho(r)$, temperature $T(r)$, and escape velocity $v_{\mathrm{esc}}(r)$ are passed without interpolation to the dark–matter module.  Capture is computed by integrating over the stellar radius the product of the local nuclear density, the dark–matter velocity distribution truncated at $v_{\mathrm{esc}}(r)$, and the differential scattering cross section; this tight coupling ensures that any structural change, no matter how rapid, feeds directly into the capture rate at the next timestep.

Pre–main–sequence growth is included through a constant user–specified accretion rate $\dot{M}_*$.
%$\dot{M}_{\mathrm{accr}}$.  
At each step \texttt{accrini} restricts the timestep to $\Delta t \lesssim 0.01\,M_*/\dot{M}_{*}$, after which \texttt{strat} inserts a new outer mass shell, shifts all thermodynamic variables inward, and mixes the accreted material with primordial (interstellar) abundances.  The Henyey solver is then called anew so that the star regains full equilibrium before the dark–matter integral is evaluated.  This procedure lets the code follow, in lock-step, how continuous mass loading alters the central density and hence modulates the accumulation of DM throughout the protostellar phase.

%-----------------------------------------------------------------
\subsection{Initial model and free parameters}
\label{sec:setup}

% ------------------------------------------------------------------
% Initial conditions and model grid
Having described the numerical machinery and its coupling to dark–matter capture, we now specify the set of Pop\,III.1 initial models to which it is applied. Each simulation is started from a chemically pristine protostellar seed of mass $M_* = 2\,M_\odot$ at an age of $9$\,yr.  The envelope composition is $X = 0.7516$ and $Y = 0.2484$ ($Z = 0$), identical to the Pop\,III values adopted by \citet{Nandal2024}.  WIMPs are injected according to the parameter vector
\begin{equation}
\begin{aligned}
[m_\chi,\langle\sigma_a v\rangle,\sigma_{\mathrm{SI}},\sigma_{\mathrm{SD}},v_\chi] &=
[100\,\mathrm{GeV},\,3\times10^{-26}\,\mathrm{cm^{3}\,s^{-1}},\\
 &\quad\,10^{-47}\,\mathrm{cm^{2}},\,10^{-41}\,\mathrm{cm^{2}},\,10\,\mathrm{km\,s^{-1}}],
\end{aligned}
\label{eq:param_set}
\end{equation}
where $m_\chi$ is the WIMP mass, $\langle\sigma_a v\rangle$ the thermally averaged annihilation cross section, and $\sigma_{\mathrm{SI}}$ and $\sigma_{\mathrm{SD}}$ the spin–independent and spin–dependent scattering cross sections, respectively.  The values are consistent with the latest constraints reported by the LZ collaboration \citep{LZ2024}.  The surrounding minihalo is modelled with two representative WIMP energy densities, $\rho_\chi = 10^{12}$ and $10^{15}\,\mathrm{GeV\,cm^{-3}}$, and a Maxwellian velocity dispersion $v_\chi = 10\,\mathrm{km\,s^{-1}}$.

To isolate the role of DM from that of mass growth, a constant gas accretion rate is prescribed by the user, taking the values $\dot{M}_* = 10^{-1},\,10^{-2},\,3\times10^{-3}, \,10^{-3}\,M_\odot\,\mathrm{yr^{-1}}$.  Combining the six background WIMP densities with the three accretion rates yields nine distinct evolutionary tracks, summarised in Table~\ref{tab:models}, along which the capture–annihilation feedback is allowed to reshape the protostar from its earliest contraction phase onward.

\begin{table*}[t]
    \centering
    \caption{Initial and final properties DM-powered stellar models.}
    \label{tab:models}
    \resizebox{\textwidth}{!}{%
    \begin{tabular}{|cc|ccccccccc|}
        \hline\hline
        $\dot{M}_*$ [$M_\odot\,\mathrm{yr}^{-1}$] &
        $\rho_{\chi}$ [$\mathrm{GeV\,cm}^{-3}$] &
        $M_{*f}$ [$M_\odot$] &
        $t_{*f}$ [yr] &
        $\Gamma_{\mathrm{Edd}}$ &
        $M_{\mathrm{cc}}$ &
        $X_{{}^{1}\mathrm{H}}$ &
        $N_{\chi,i}$ &
        $M_{\chi,i}$ [$M_\odot$] &
        $N_{\chi,f}$ &
        $M_{\chi,f}$ [$M_\odot$] \\
        \hline
        $3\cdot10^{-3}$ & 0
        & 436
        & $1.449\cdot10^{5}$
        & 0.6889
        & 0.9191
        & 0.7341
        & 0
        & 0
        & 0
        & 0 \\

        $3\cdot10^{-3}$ & $10^{12}$
        & 443
        & $1.471\cdot10^{5}$
        & 0.6898
        & 0.9159
        & 0.7352
        & $1.7121\cdot10^{47}$
        & $1.5349\cdot10^{-8}$
        & $9.951\cdot10^{49}$
        & $8.921\cdot10^{-6}$ \\

        $3\cdot10^{-3}$ & $10^{13}$
        & 445
        & $1.483\cdot10^{5}$
        & 0.6886
        & 0.9344
        & 0.7445
        & $5.4141\cdot10^{47}$
        & $4.8537\cdot10^{-8}$
        & $3.165\cdot10^{50}$
        & $2.837\cdot10^{-5}$ \\

        $3\cdot10^{-3}$ & $10^{14}$
        & 702
        & $2.336\cdot10^{5}$
        & 0.7088
        & 0.9701
        & 0.7516
        & $1.7121\cdot10^{48}$
        & $1.5349\cdot10^{-7}$
        & $1.894\cdot10^{51}$
        & $1.698\cdot10^{-4}$ \\

        $3\cdot10^{-3}$ & $5\cdot10^{14}$
        & 429209
        & $1.431\cdot10^{8}$
        & 1.3109
        & 1.0000
        & 0.7515
        & $3.146\cdot10^{48}$
        & $2.8204\cdot10^{-7}$
        & $3.489\cdot10^{55}$
        & 3.128 \\

        $3\cdot10^{-3}$ & $10^{15}$
        & 516575
        & $1.722\cdot10^{8}$
        & 2.1133
        & 1.0000
        & 0.7516
        & $5.4141\cdot10^{48}$
        & $4.8537\cdot10^{-7}$
        & $8.27\cdot10^{55}$
        & 7.414 \\

        $3\cdot10^{-3}$ & $10^{16}$
        & 50818
        & $1.694\cdot10^{8}$
        & 4.1771
        & 1.0000
        & 0.7516
        & $1.7121\cdot10^{49}$
        & $1.5349\cdot10^{-6}$
        & $4.403\cdot10^{55}$
        & 3.947 \\

        $10^{-2}$ & $10^{13}$
        & 364980
        & $3.6498\cdot10^{7}$
        & 0.9922
        & 0.9990
        & 0.7515
        & $5.4141\cdot10^{47}$
        & $4.8537\cdot10^{-8}$
        & $6.0043\cdot10^{54}$
        & 0.5383 \\

        $10^{-2}$ & $10^{15}$
        & 106440
        & $1.0644\cdot10^{7}$
        & 1.2548
        & 1.0000
        & 0.7516
        & $5.4141\cdot10^{48}$
        & $4.8537\cdot10^{-7}$
        & $5.1255\cdot10^{54}$
        & 0.4595 \\

        $10^{-3}$ & $10^{15}$
        & 139160
        & $1.3916\cdot10^{8}$
        & 1.3704
        & 1.0000
        & 0.7516
        & $5.4141\cdot10^{48}$
        & $4.8537\cdot10^{-7}$
        & $1.516\cdot10^{55}$
        & 1.359 \\

        $10^{-1}$ & $10^{15}$
        & 140020
        & $1.4002\cdot10^{6}$
        & 1.3874
        & 1.0000
        & 0.7516
        & $5.4141\cdot10^{48}$
        & $4.8537\cdot10^{-7}$
        & $1.392\cdot10^{55}$
        & 1.248 \\
        \hline
    \end{tabular}%
    }
    \tablefoot{The first two columns give the prescribed gas accretion rate, $\dot{M}_*$, and the ambient WIMP density, $\rho_{\chi}$. The remaining columns give the values at the end of each calculation: the final stellar mass, $M_{*f}$; final stellar age, $t_{*f}$; Eddington factor, $\Gamma_{\mathrm{Edd}}$; final convective-core mass fraction, $M_{\mathrm{cc}}$; final central hydrogen mass fraction, $X_{{}^{1}\mathrm{H}}$; final WIMP number, $N_{\chi,f}$; and final WIMP mass, $M_{\chi,f}$. The initial WIMP number and mass are denoted by $N_{\chi,i}$ and $M_{\chi,i}$, respectively. The WIMP masses $M_{\chi,i}$ and $M_{\chi,f}$ are given in units of $M_\odot$.}
\end{table*}

%-----------------------------------------------------------------
\subsection{Radiative feedback}

To model radiative feedback limiting the growth of our DM-powered stars, we implement two key processes: Eddington-limited accretion and photoevaporative mass loss. Accretion is suppressed when the stellar luminosity approaches the Eddington limit, defined by $L_{\text{Edd}} = \frac{4\pi G M_* c}{\kappa}$, where $\kappa \approx 0.34 \, \text{cm}^2 \, \text{g}^{-1}$ for electron scattering in primordial gas. Accretion is limited such that $\dot{M}_* \lesssim \frac{L_{\text{Edd}} R_*}{G M_*}$, where $R_*$ and $M_*$ are the stellar radius and mass.

We also include photo-evaporative feedback driven by ionising UV radiation from the growing protostar. This feedback acts by photo-evaporating the circumstellar accretion disk, thereby suppressing the supply of gas to the star rather than directly removing mass from the stellar surface. The resulting disk photo-evaporation rate is estimated as (see equation 67 of) \citep{McKeeTan2008}
\begin{equation}
\dot{M}_{\mathrm{pe}} \approx 4.1 \times 10^{-5}
S_{49}^{1/2}
\left( \frac{T_i}{10^{4}\,\mathrm{K}} \right)^{0.4}
\left( \frac{M_*}{100\,M_\odot} \right)^{1/2}
\,M_\odot\,\mathrm{yr}^{-1},
\label{eq:mdot_pe}
\end{equation}
where $S_{49}$ is the ionising photon rate in units of $10^{49} \, \text{s}^{-1}$ and $T_i$ is the ionised gas temperature. Accretion is terminated once the disk photoevaporation rate exceeds the mass accretion rate through the disk, $\dot{M}_{\text{pe}} \geq \dot{M}_*$. This combined feedback sets a natural limit on the final mass of DM-powered protostars.

\section{Results}\label{Sec:Results}
\subsection{DM annihilation and stellar structure}\label{Sec:Transport}

All models begin their evolution as nearly fully convective $2M_\odot$ protostellar seeds. The choice of initial baryonic and DM parameters strongly affect the stellar structure and its subsequent evolution. We begin this section by exploring the effect of changing the background DM density ($\rho_{\chi}$) whilst keeping the WIMP mass (100 GeV) and stellar accretion rate (3$\times$10$^{-3}\:M_\odot\:$yr$^{-1}$) constant. The capture of dark-matter and accretion of baryonic matter commences once the age of the models is nine years. This is done to ensure stable numerical convergence for the initial structures. We now break down the evolution of the five supermassive models in four different stages, as shown in the left and right panels of Figure~\ref{fig:1_HR_and_tcrhoc}. In addition to the 6 DM-powered models, we also computed a case with no DM reservoir or capture; this case represents a classical baryonic Pop~III star. 

\subsubsection{Stage I: Initial contraction} 
\begin{figure*}[h]
    \centering
    \includegraphics[width=0.52\textwidth]{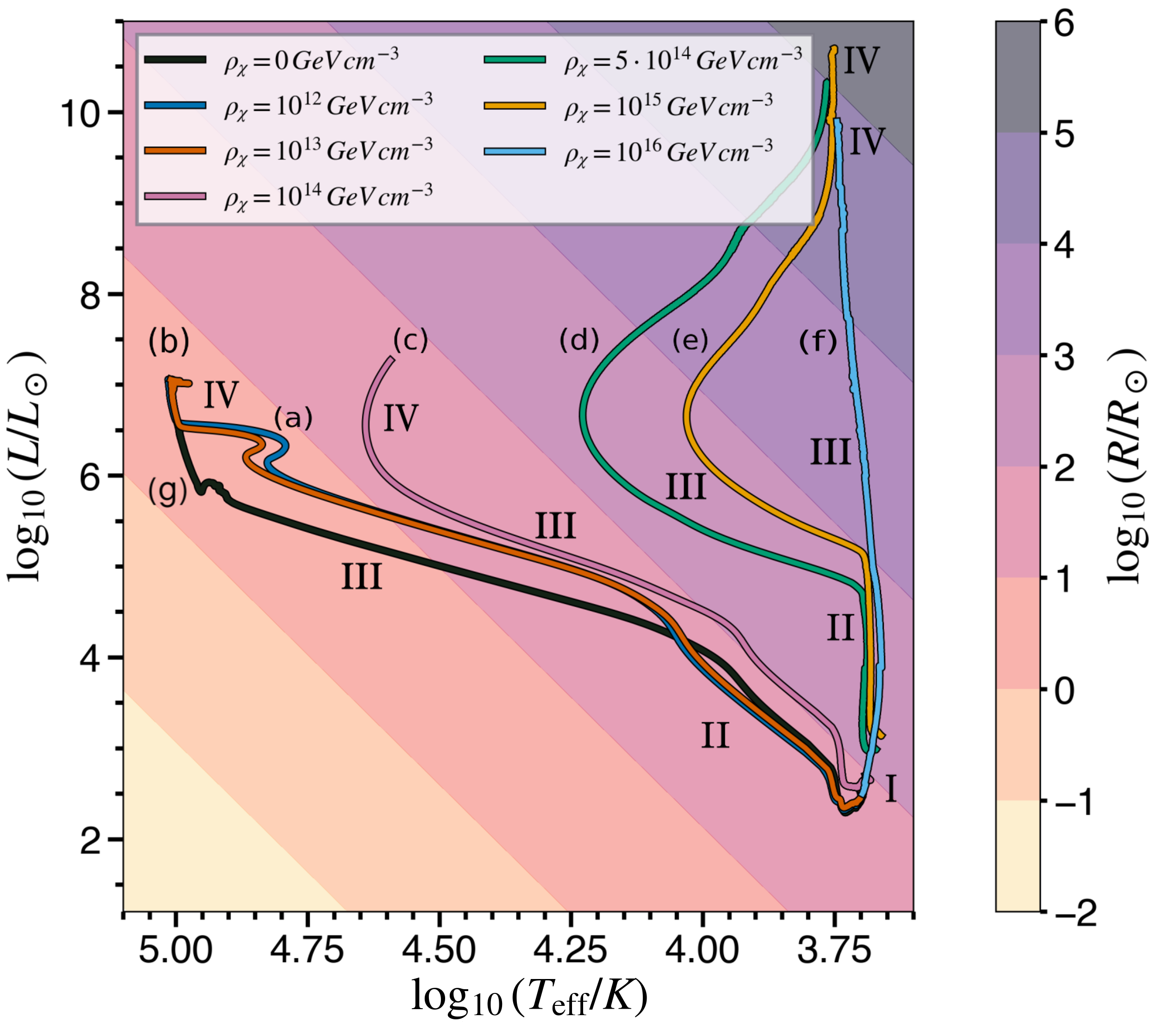}
    \includegraphics[width=0.46\textwidth]{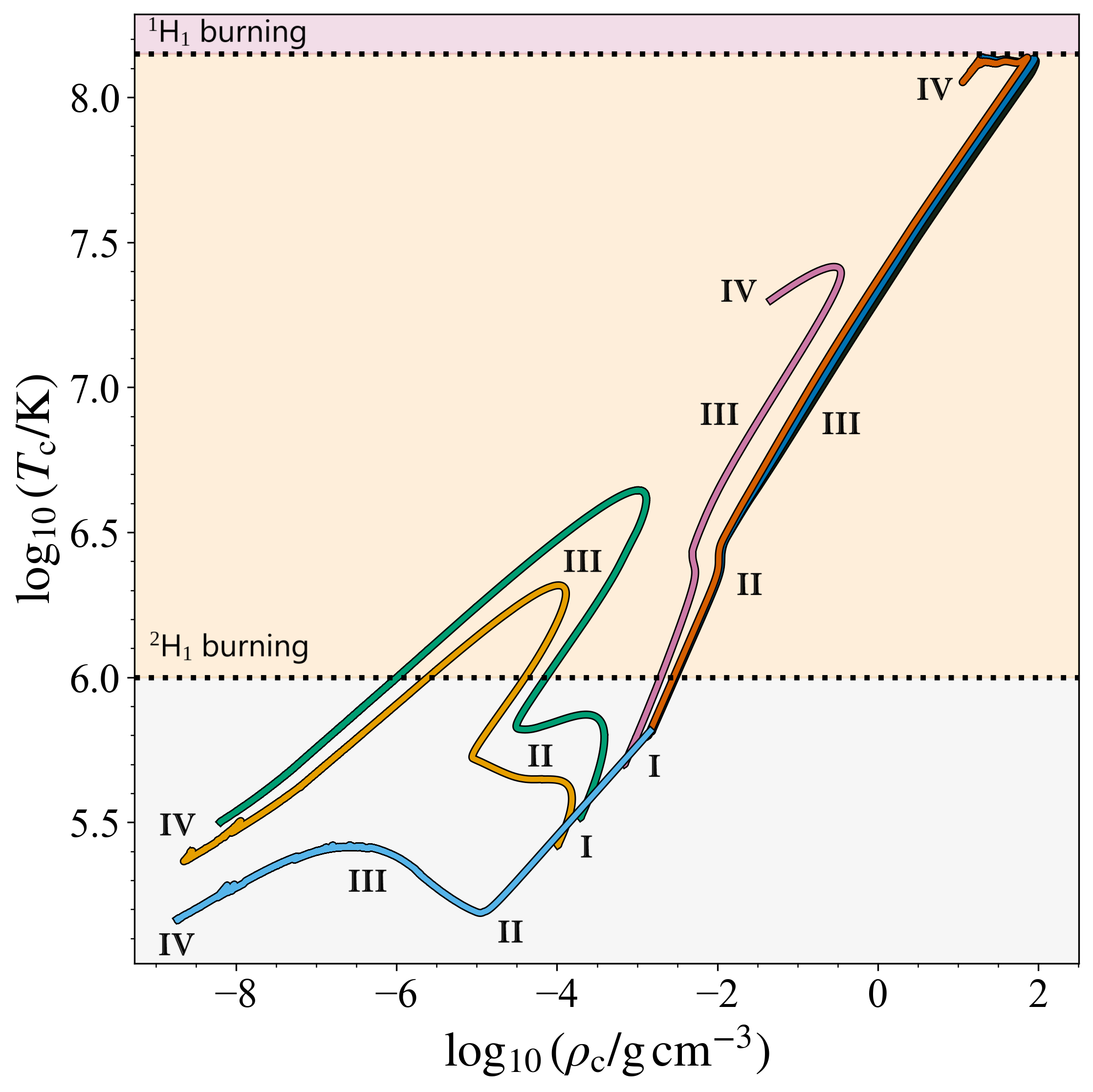}
    \caption{Seven massive and supermassive stellar models at WIMP densities ranging from 10$^{12}$ to 10$^{16}$ GeV cm$^{-3}$ at a constant accretion rate of 3$\times$10$^{-3} M_\odot\:$yr$^{-1}$ labelled from (a) to (f) respectively. Model (g) represents the case of standard Pop III star formation without any WIMP capture or annihilation. \textit{(a) Left:} Hertzsprung-Russell (HR) diagram with isoradii depicted using a colourbar. \textit{(b) Right:} Evolution of central temperature versus central density. The grey, yellow, and pink coloured zones depict no nuclear burning, deuterium burning, and hydrogen burning, respectively. The Roman numerals I–IV indicate successive evolutionary stages: (I) pre-main-sequence (pre-MS), (II) contraction phase, (III) nuclear-burning phase, and (IV) the onset of instability.}
    \label{fig:1_HR_and_tcrhoc}
\end{figure*}

This is the start of the pre-main sequence evolution where the models are separated into two groups; models (a), (b), (c), and (g) have an effective temperature, log~($T_{\rm eff}$) = 3.7 and luminosity, log~($L/L_\odot$) = 2.75, whereas models (e) and (f) have an effective temperature, log~ ($T_{\rm eff}$) = 3.65 and luminosity, log~($L/L_\odot$) = 3.10 (see left panel of Figure~\ref{fig:1_HR_and_tcrhoc}). The difference in the starting positions is due to the differences in the initial structure, dictated by the DM reservoir of each model. The DM reservoir is dependent on the background WIMP density ($\rho_\chi$), and this reservoir exists due to an evolutionary phase that led to the formation of these $2M_\odot$ seeds. The effect of increasing DM reservoirs is also evident in the inner regions of all models, where the central temperatures and densities are inversely related ($\rho_c, T_c \propto 1/\rho_{\chi}$), as seen in the right panel of Figure~\ref{fig:1_HR_and_tcrhoc}. All models undergo an initial contraction phase that lasts for 20 - 500 years, and this duration is inversely proportional to the background DM density ($\rho_\chi$). In other words, model (a) spends the first 500 years contracting whereas model (f) contracts for only 20 years. 

\subsubsection{Stage II: Dark matter annihilation and luminosity budgets} The choice of background WIMP densities and the effects of WIMP annihilation become apparent at this stage. With accretion rate constant, models (a), (b), (c), and (g) continue to contract while models (d), (e), and (f) expand in radius almost vertically along the Hayashi line. This effect can be better understood by looking at the luminosity budgets of each model, as shown in Figure~\ref{fig:2_lumbudget}. At an age of 500 years, we find that the largest source of energy for models (a), (b), and (c) comes from the gravitational contraction, followed by nuclear fusion of Deuterium. The energy generated from WIMP annihilation remains a factors of a few below the energy released from gravitational contraction, even for model (c). In models (d) and (e) with higher background WIMP density, the largest source of energy is instead WIMP annihilation, followed by gravitational contraction. 

\subsubsection{Stage III: Luminosity wave and dark matter heating} 

\begin{figure*}[!]
    \centering
    \begin{minipage}{0.33\textwidth}
        \centering
        \includegraphics[width=\linewidth]{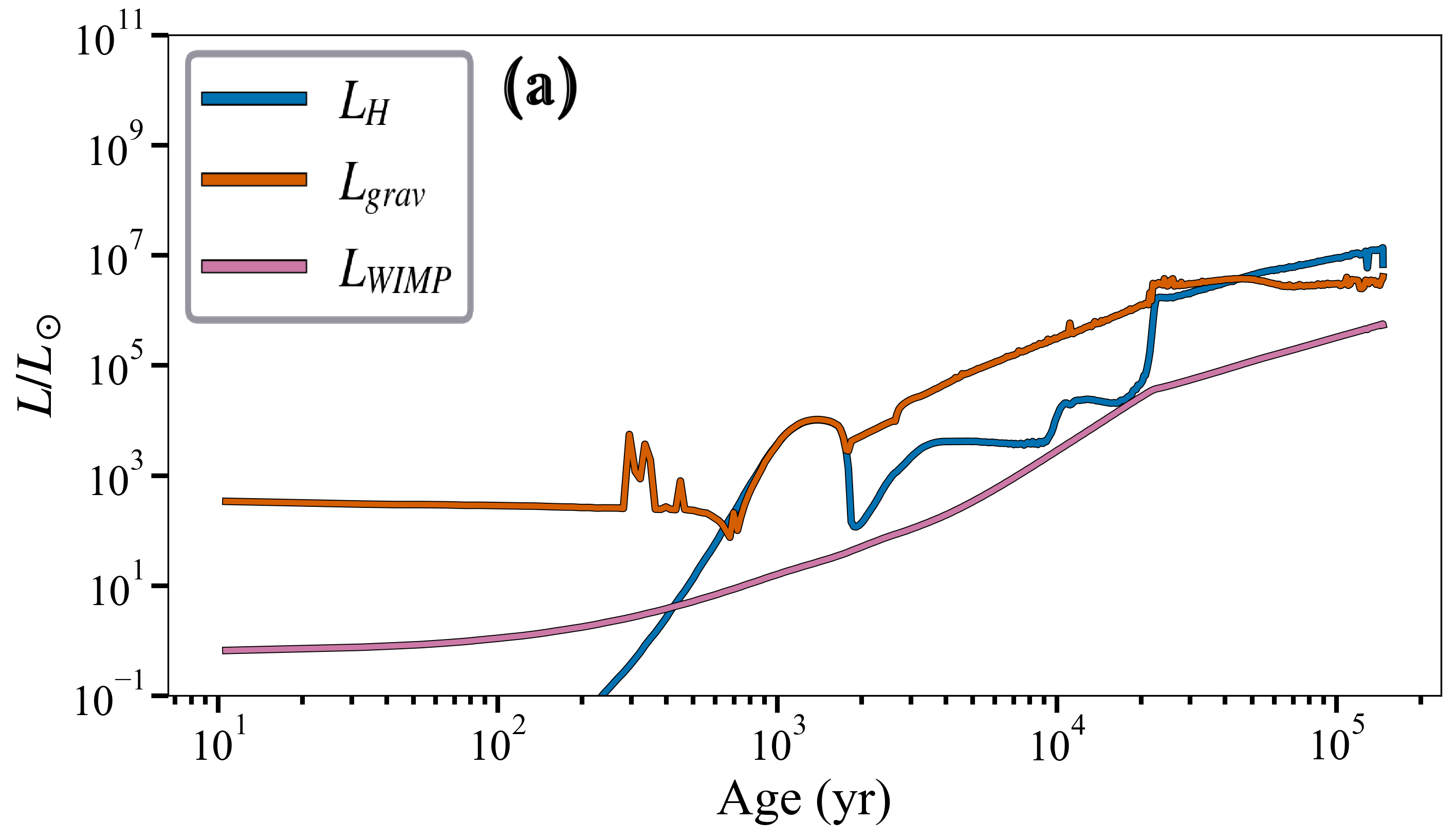}
    \end{minipage}%
    \begin{minipage}{0.33\textwidth}
        \centering
        \includegraphics[width=\linewidth]{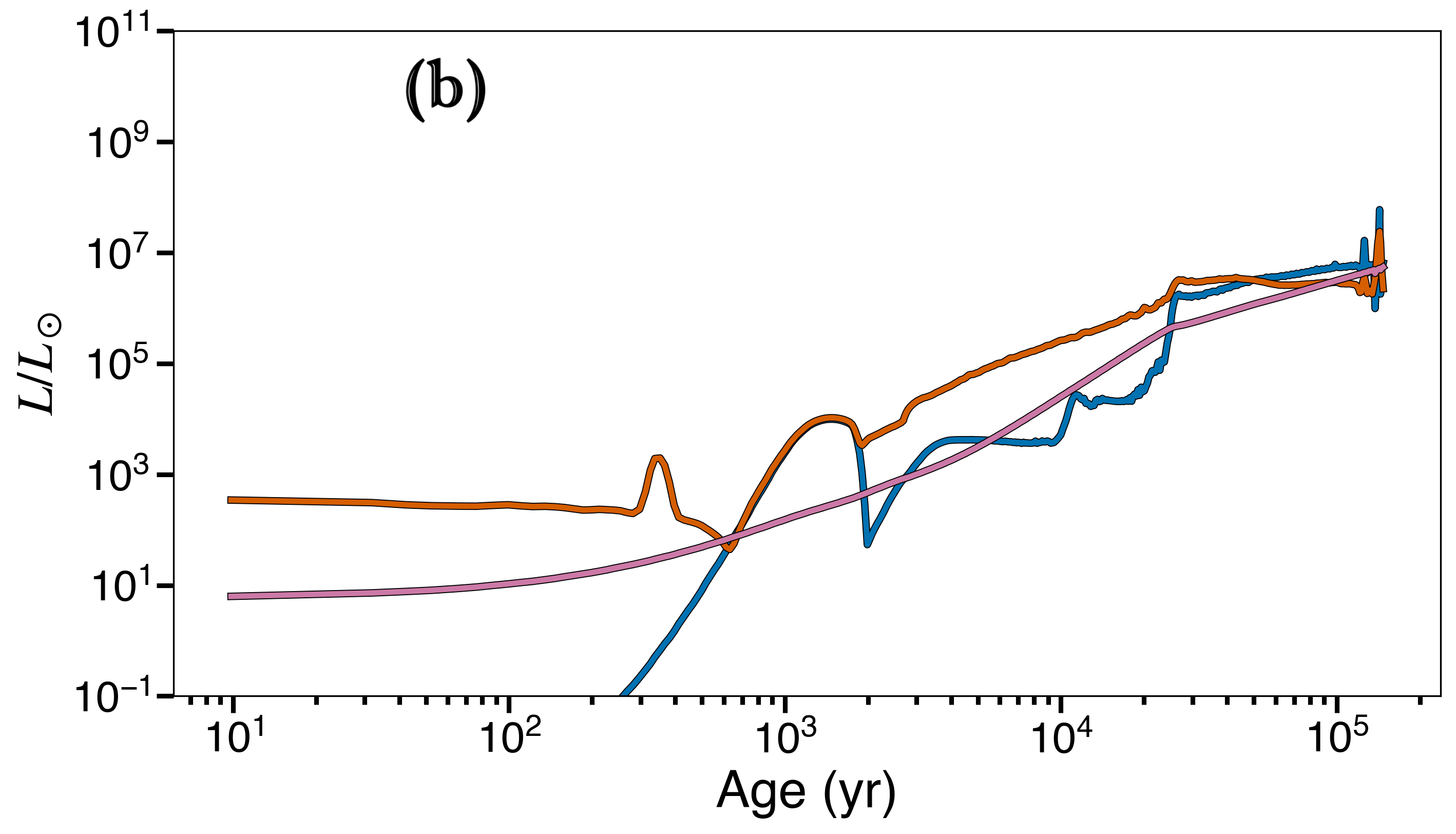}
    \end{minipage}%
    \begin{minipage}{0.33\textwidth}
        \centering
        \includegraphics[width=\linewidth]{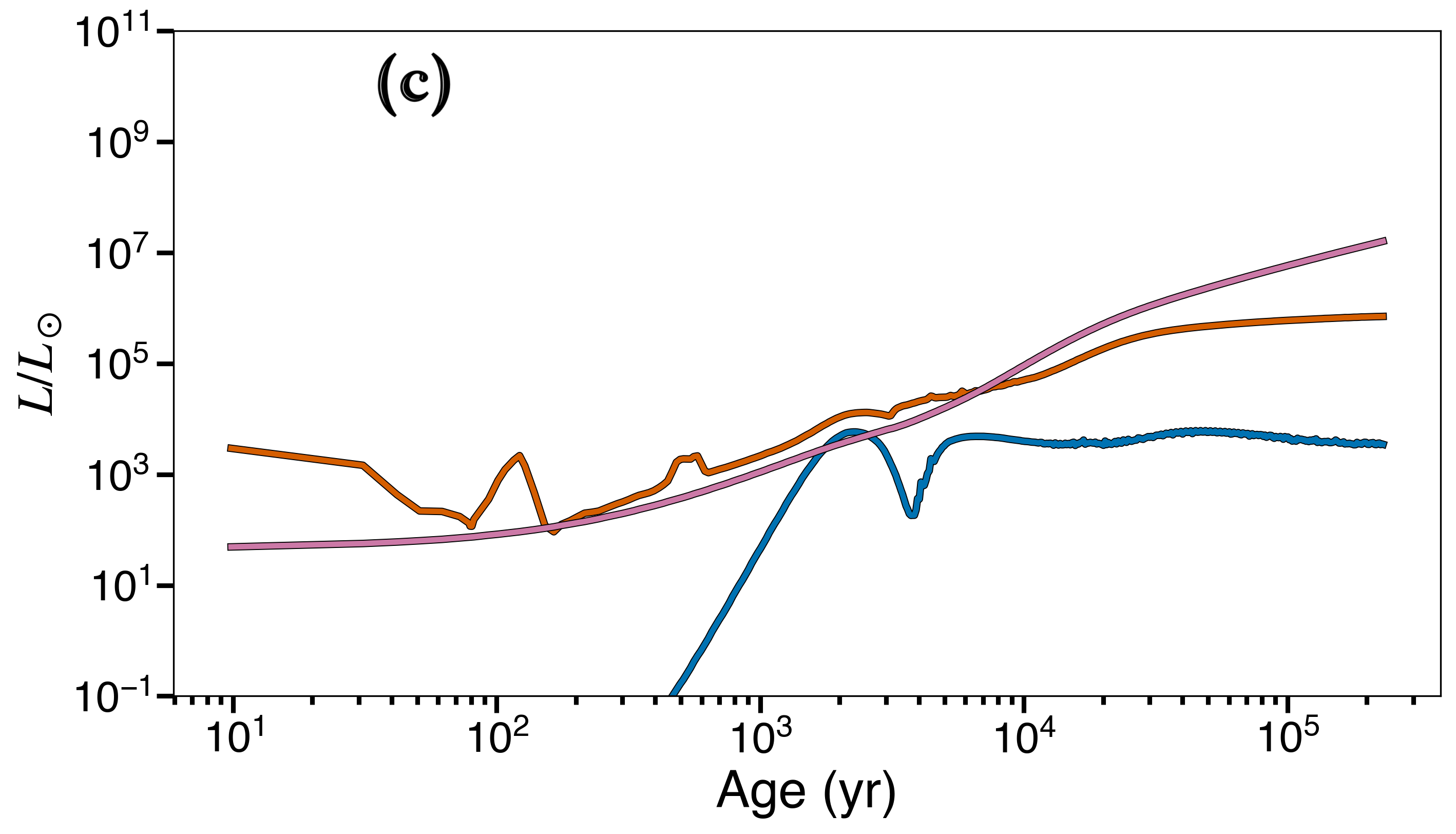}
    \end{minipage}

    \vspace{0.5cm}  % Adds vertical space between rows

    \begin{minipage}{0.33\textwidth}
        \centering
        \includegraphics[width=\linewidth]{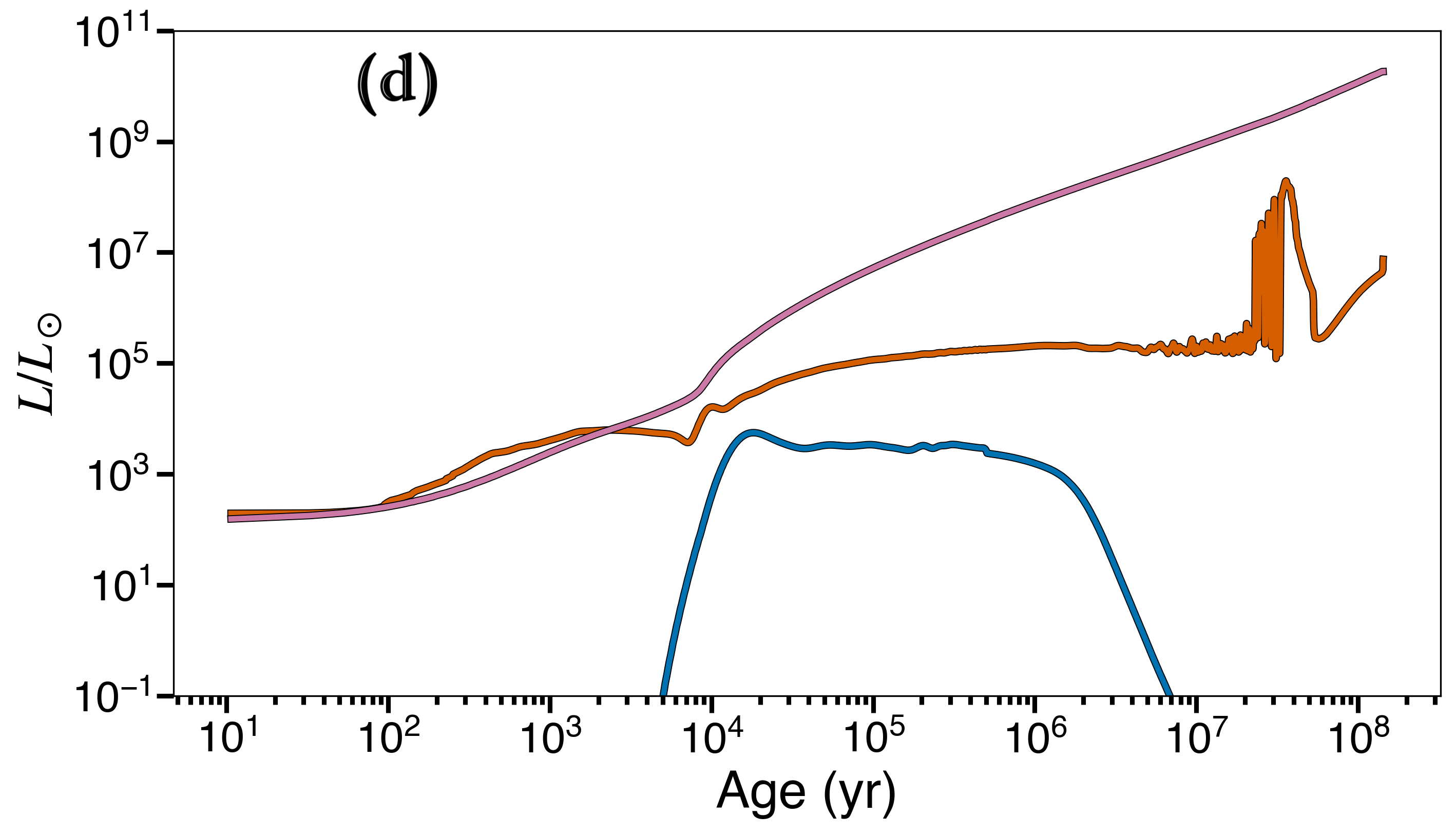}
    \end{minipage}%
    \begin{minipage}{0.33\textwidth}
        \centering
        \includegraphics[width=\linewidth]{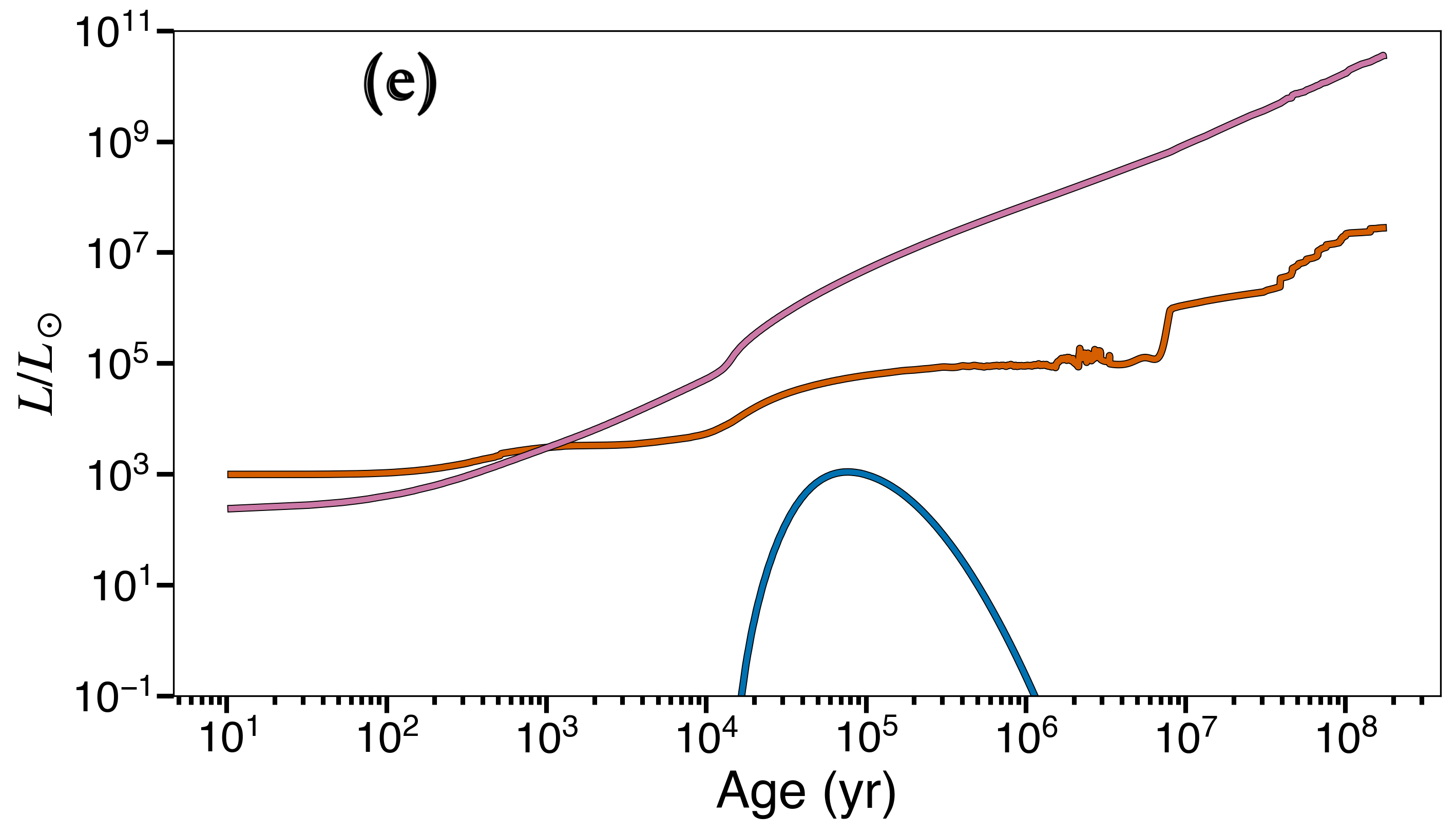}
    \end{minipage}%
    \begin{minipage}{0.33\textwidth}
        \centering
        \includegraphics[width=\linewidth]{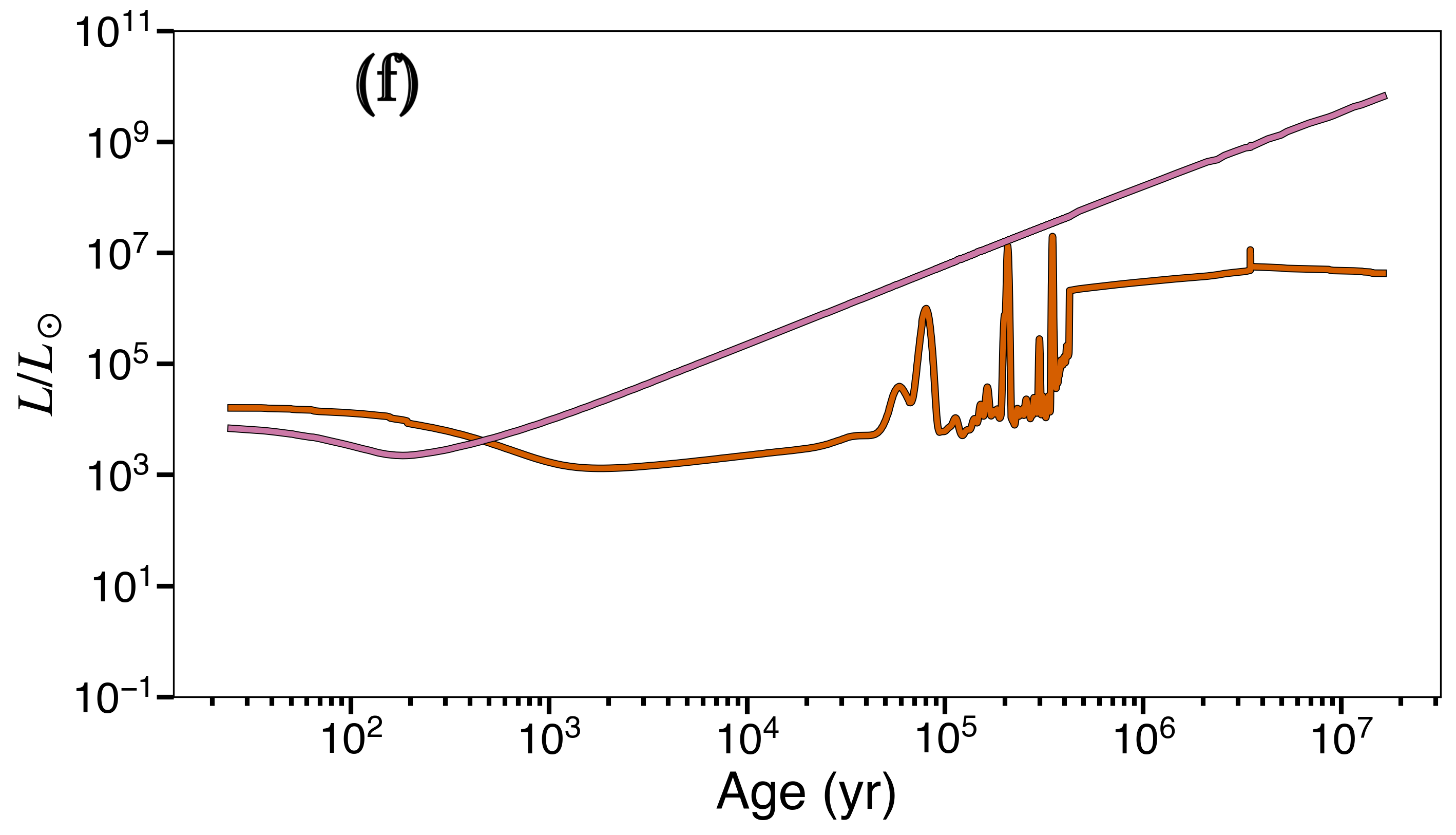}
    \end{minipage}

    \vspace{0.5cm}

    \begin{minipage}{0.33\textwidth}
        \centering
        \includegraphics[width=\linewidth]{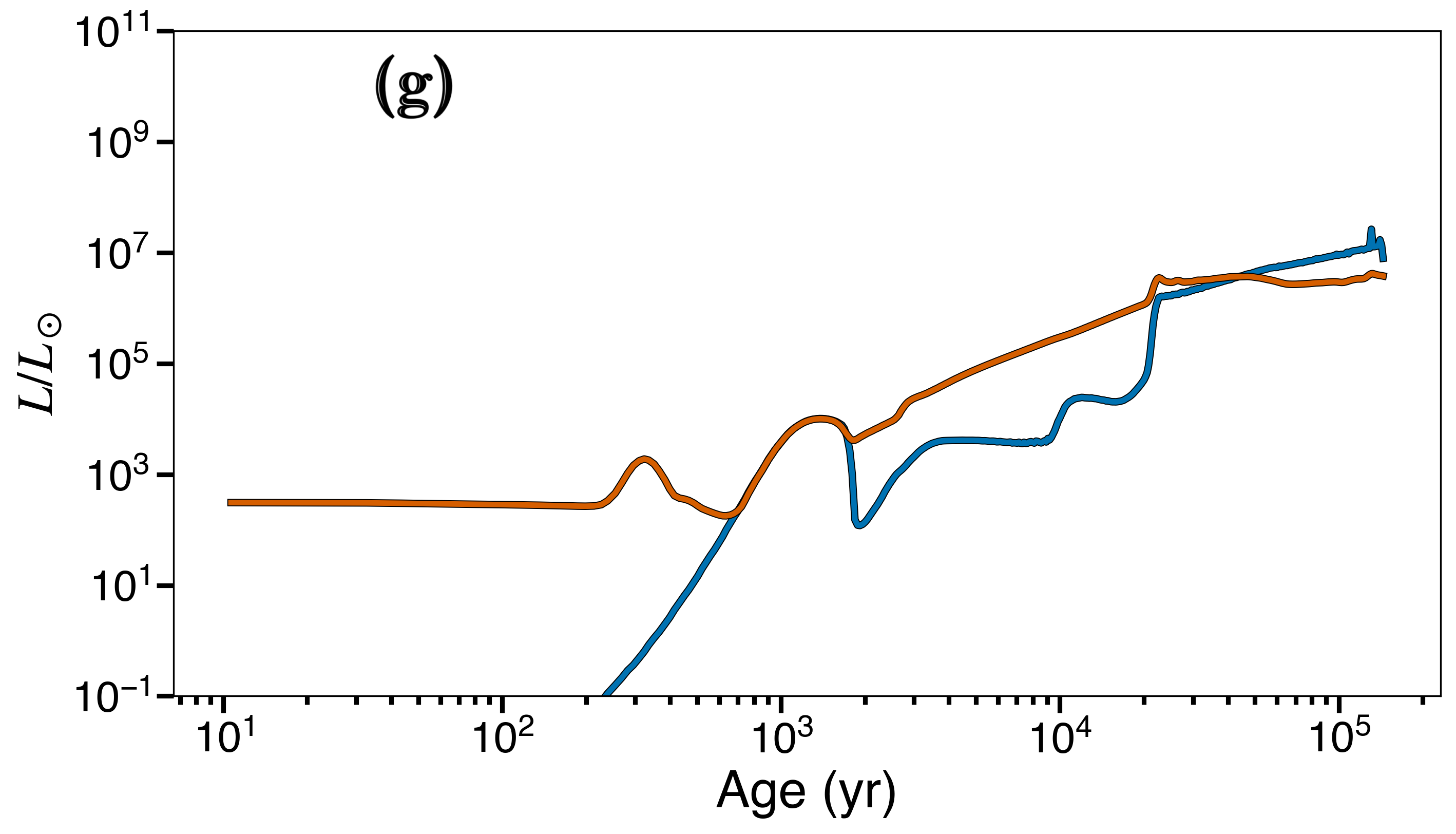}
    \end{minipage}

    \caption{Instantaneous luminosity budgets of supermassive protostars accreting at 
    $\dot{M}_* = 3 \times 10^{-3}\,M_\odot\ \mathrm{yr^{-1}}$ and evolving in DM halos of constant WIMP density. Panels (a)–(f) correspond to $\rho_{\chi}=10^{12},10^{13},10^{14},5\times10^{14},10^{15},10^{16}\ \mathrm{GeV\ cm^{-3}}$, 
    respectively; the seventh panel (g) is a model without any WIMP capture or annihilation.}
    \label{fig:2_lumbudget}
\end{figure*}

Once all models reach an age $\geq$ 500 years, they encounter luminosity wave episodes \citep{Larson1972}. The continual accretion of matter onto the stellar surface causes these models to contract and increase the central temperature and density (see right panel of Figure~\ref{fig:1_HR_and_tcrhoc}). This changes the central opacity, which leads to an increase in the luminosity of the models. This luminosity may only migrate outwards from the centre and once it breaks at the surface, it produces a luminosity wave. Previous works have shown that the choice of accretion rate at this stage is crucial as it determines whether a model contracts towards the ZAMS or expands towards the Hayashi limit \citet{Hosokawa_2010, Nandal2024, Herrington2023}. The accretion rate of 3$\times$10$^{-3} M_\odot$yr$^{-1}$ is below the critical accretion rate of 2.5$\times$ 10$^{-2} M_\odot$yr$^{-1}$ \citep{Nandal2024}, which forces models (a), (b), and (c)) to migrate to the blue side of HR diagram (left panel of Figure~\ref{fig:1_HR_and_tcrhoc}). This effect can be understood by comparing the luminosity budgets of these models (Figure~\ref{fig:2_lumbudget}) once they reach an age of 1000 years. The evolution  of models (a) and (b) is being dominated by gravitational contraction, followed by nuclear burning and WIMP annihilation. In case of model (c), gravitational contraction still dominates the luminosity budget; however, the second largest contribution comes from WIMP annihilation instead of nuclear burning. Model (c) is the transitory model beyond which the effect of WIMP annihilation on the stellar structure begins to dominate over nuclear burning beyond 1000 years. The effects of WIMP annihilation on the stellar structure are at the forefront of total luminosity budget in model (d), where gravitational contraction and nuclear burning become the second and third highest luminosity sources. Consequently, the transition of model (d) towards the ZAMS is halted, and the model instead prepares for an expansion towards the Hayashi line. Model (e) also undergoes a similar transition as model (d), but in addition maintains a larger stellar radius due to an increased WIMP annihilation rate from the beginning of the computation. Finally, model (f) does not undergo any contraction since its luminosity budget is dominated by DM annihilation right from stage I, and instead climbs vertically along the Hayashi line. In addition to the models powered by WIMP annihilation, model (g), which lacks any DM capture (see Figures~\ref{fig:1_HR_and_tcrhoc}, \ref{fig:2_lumbudget}), has a straightforward contraction phase towards the ZAMS. With gravitational contraction powering model (g), it reaches the ZAMS at a mass of 101~$M_\odot$ over a time of 1.5$\times$10$^{5}$ years. 

\subsubsection{Stage IV: Nuclear burning versus dark matter annihilation}

The effects of DM annihilation on the stellar structure become the most pronounced once the models reach an age of about 10$^5$ years. Models (a) and (b), undergo a short expansion phase at a luminosity log~($L/L_\odot$) = 6.25 and effective temperature, log~($T_{\rm eff}$) = 4.82. This is due to the combined effect of the onset of core hydrogen burning and WIMP annihilation. In the case of model (a) with lower WIMP annihilation rate, core hydrogen burning sets in sooner as a consequence of rising central temperature and density, which in turn is governed by gravitational contraction (see right panel of Figure~\ref{fig:1_HR_and_tcrhoc}). The excess energy generated does work in expanding the envelope, but with gravitational energy still being the dominant source of energy as the model contracts to the ZAMS at a mass of 102~$M_\odot$. The effects of radiative feedback become significant during this stage, and as the model reaches the ZAMS, mass loss rates due to photoevaporation become equivalent to the accretion rate (see Equation \ref{eq:mdot_pe}. The condition for radiative feedback is satisfied and the final mass of the model is 443~$M_\odot$. 

Model (b) also undergoes a short expansion phase but with a higher WIMP annihilation rate than model (a), the energy released from nuclear burning begins to plateau and subsequently diminishes (see panel (b) in Figure~\ref{fig:2_lumbudget}). This results in a shorter redward excursion and, with gravitation contraction still dominating the luminosity budget, the model eventually contracts to the ZAMS at a mass of 108~$M_\odot$, and accretion terminates due to radiative feedback at a final mass of 445~$M_\odot$. 

In model (c), the effects of WIMP annihilation become clearly evident as it never contracts to the ZAMS throughout its evolution. However, upon considering the effects of radiative feedback, we find that once the mass of the model reaches 702~$M_\odot$, radiative losses dominate over accretion and the model attains its final mass. The evolution beyond this point is not followed but the model is expected to contract to the ZAMS over the thermal timescale. We note that while radiative feedback from the star can terminate baryonic accretion from the disk, it does not directly halt DM accretion from the surrounding halo. Continued DM supply could, in principle, extend the period during which WIMP annihilation contributes to the stellar luminosity, potentially delaying contraction to the ZAMS and modifying the subsequent thermal and structural evolution of the dark star.

Models (d), (e), and (f) avoid any radiative losses due to their larger stellar radii ($> 200 R_\odot$) and cooler surface temperatures (log~($T_{\rm eff}/{\rm K}$) < 4.25), as a consequence of the high WIMP annihilation rates. They continue to accrete mass and are fully supported by WIMPs until they reach a mass of 10$^{5}\:M_\odot$, at which point the computation is terminated (Figure~\ref{fig:2_lumbudget}). Their final fates are discussed in an upcoming section. These models can be characterised as DM-powered protostars models whose formation and subsequent stellar evolution do not depend on nuclear burning stages, but instead on DM heating powering their protostellar cores. The left and right panels of Figure~\ref{fig:1_HR_and_tcrhoc} show clear evolutionary pathways and central conditions that distinguish ``standard'' Pop III protostars with no influence of WIMP annihilation from the DM-powered protostars.  

\subsection{Ionising photon production and radiative feedback}\label{Sec:Ionising}

\begin{figure*}[h]
    \centering
    \includegraphics[height=7cm]{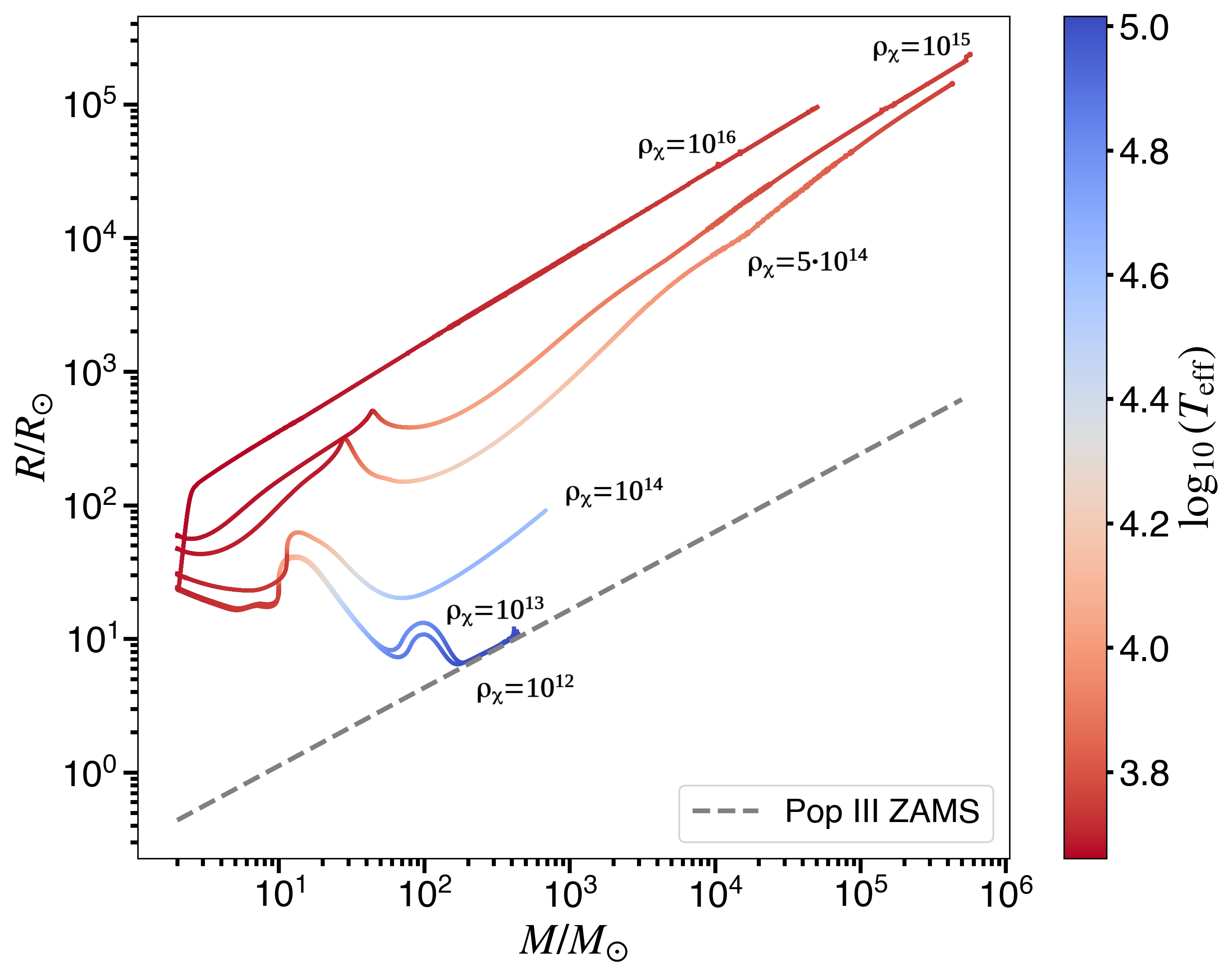}
    \includegraphics[height=7cm]{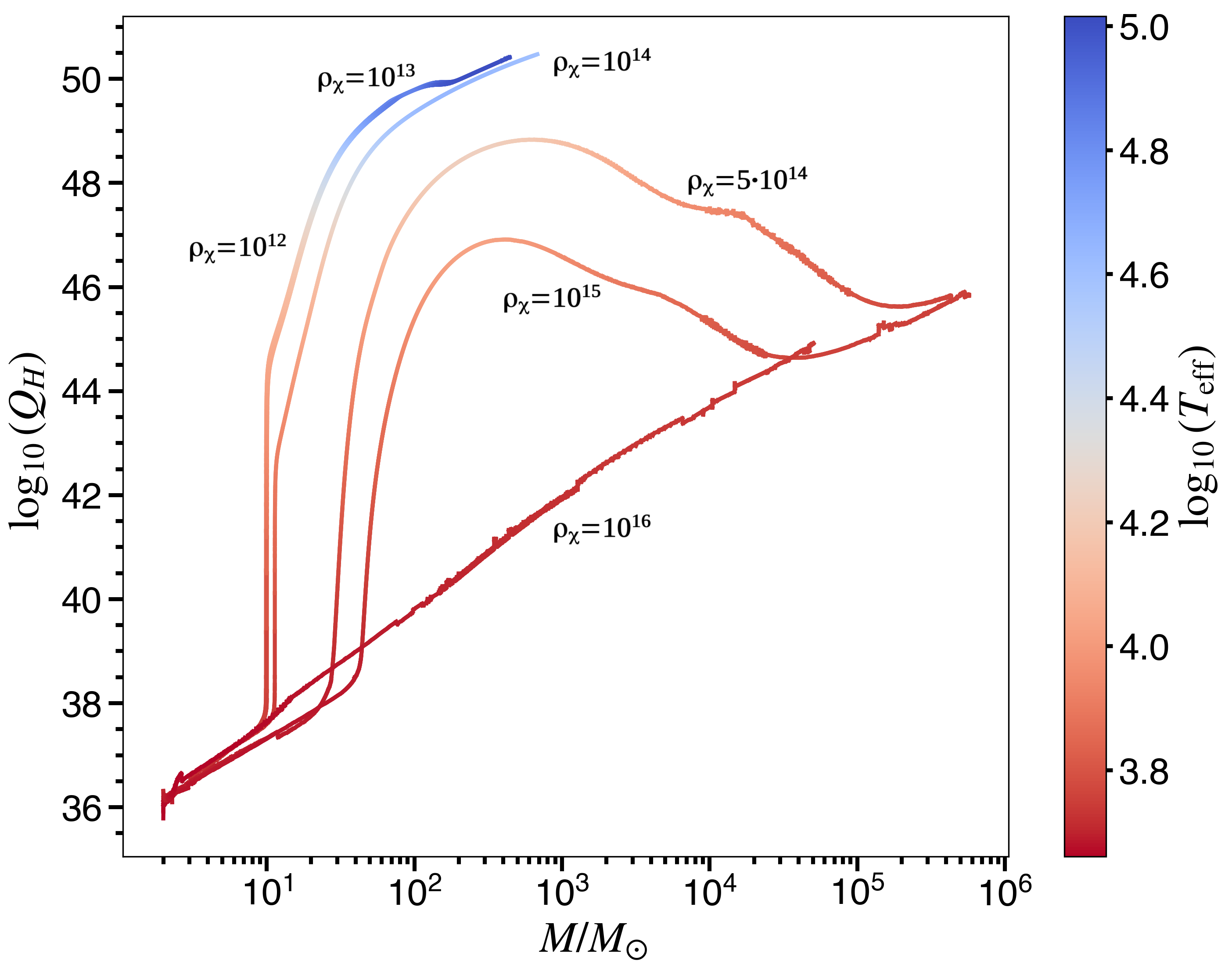}
    \caption{Hydrogen ionising photon production and structural evolution of accreting Pop~III protostars at a fixed accretion rate of $\dot{M}_*=3\times 10^{-3}\,M_\odot\,\mathrm{yr}^{-1}$ for five background WIMP densities ($\rho_\chi = 10^{12}$--$10^{16}\,\mathrm{GeV\,cm^{-3}}$). \textit{(a) Left:} Hydrogen-ionising photon rate $Q_H$ versus stellar mass, colour-coded by $\log_{10}(T_{\mathrm{eff}}/{\rm K})$. \textit{(b) Right:} Stellar radius as a function of mass for the same models and colour scale, highlighting how stronger DM annihilation heating suppresses Kelvin-Helmholtz contraction and keeps $T_{\mathrm{eff}}$, and hence $Q_H$ low.}
    \label{fig:QH_R}
\end{figure*}

Having established in Sect.~\ref{Sec:Transport} how DM annihilation heating modifies the global structure of our supermassive protostars, we now turn to the effects on stellar radii, corresponding output of hydrogen–ionising photons, $Q_H$, and its implications for radiative feedback. Figure~\ref{fig:QH_R} (left) depicts the evolution of stellar radius versus mass, colour coded by the surface temperature; the right‐hand panel shows $Q_H$ as a function of stellar mass for the five models.  Both panels employ the same ordering in ambient DM density, increasing from $\rho_\chi = 10^{12}$ to $10^{16}\,\mathrm{GeV\,cm^{-3}}$.

The ionising luminosity is initially negligible ($\log Q_H\sim36$) because all seeds are bloated and cool (stellar radius greater than 20 R$_\odot$).  Once the low‐DM models ($\rho_\chi =10^{12}$--$10^{14}$\,GeV\,cm$^{-3}$) finish their Kelvin–Helmholtz (KH) contraction phase at $M_*\sim10$--$30\,M_\odot$, their effective temperatures climb above $4\times10^{4}$\,K (blue/yellow colours) and $Q_H$ jumps by ten orders of magnitude to $\log Q_H\simeq45$--49. In contrast, the high‐DM models ($\rho_\chi = 10^{16}$) never attain $T_{\rm eff}\gtrsim3\times10^{4}$\,K; DM annihilation energy balances radiative losses before significant contraction can occur, so the envelope remains puffed up ($R\sim10^{3}$--$10^{4}R_\odot$, see left panel of Figure~\ref{fig:QH_R}) and $Q_H$ climbs only gradually to $\log Q_H\lesssim45$ even at $M_*\sim10^{5}M_\odot$ (see right panel of Figure~\ref{fig:QH_R}.

The ($\rho_\chi =10^{15}$) case behaves in‐between: partial contraction raises $T_{\rm eff}$ enough for stellar radius to be $\approx$ 100 R$_\odot$ and consequently, log$Q_H$ to reach 37. Once the model reaches a mass of 600 M$_\odot$, following a short contraction phase where stellar radius decreases from 400 - 300 R$_\odot$, the model continues to expand in radius as it accretes to a higher mass. This monotonic increase can be attributed to the effects of DM annihilation on the stellar structure that becomes dominant once it reaches a mass of 600 M$_\odot$. The large stellar radius dictates the log$Q_H$ values and we find that during the protostellar growth phase, the log$Q_H$ never exceeds 47.

Finally the $\rho_\chi =10^{16}$ case maintains the largest stellar radius of all models throughout the evolution. Consequently, the log$Q_H$ values remain the lowest until the model reaches its final mass of 50,818 M$_\odot$ (the final mass here is dictated by numerical convergence issues). 

Another way to describe the photon production is to relate the ionising production rate to the effective temperature as 
\[
   Q_H \;\propto\; \frac{L}{k T_{\rm eff}}\,
        \exp\!\Bigl[-\tfrac{13.6\,\mathrm{eV}}{kT_{\rm eff}}\Bigr],
\]
so even modest differences in $T_{\rm eff}$ on the Wien tail produce exponential changes in $Q_H$.  Since DM annihilation heating acts foremost by delaying KH contraction, it regulates $T_{\rm eff}$ and therefore the strength of radiative feedback.  Stars with weaker DM influence evolve along compact, hot tracks and emit copious ionising radiation that can ionise (and perhaps evacuate) their natal clouds.  Conversely, the most DM‐dominated objects remain cool protostars, whose feedback is effectively quenched despite their enormous masses.

The trends in Fig.~\ref{fig:QH_R} thus demonstrate a clear dichotomy; Only when the protostar can contract and heat its surface does it become an efficient source of hydrogen‐ionising photons.  Strong DM heating limits this contraction and keeps $Q_H$ orders of magnitude lower, implying that such objects may grow unimpeded to $\gtrsim10^{5}M_\odot$ without significant H \textsc{ii} regions or radiative barriers. It is due to this property and the general evolutionary trends described in previous section, we consider model with $\rho_\chi = 10^{15}$ to be our fiducial candidate for a DM-powered SMS scenario. 

%------------------------------------------------------------------

%------------------------------------------------------------------

%------------------------------------------------------------------
\subsection{General relativistic stability and the impact of DM}
\label{Sec:GR}

General relativistic instability (GRI) has been extensively studied for standard supermassive stars in the mass range of 10,000 - 1,000,000 $M_\odot$ \citep{Chandrasekhar_1964b,Hosokawa_2010, Lionel2018, Nagele2023, Nandal2024c}. To establish whether DM-powered stars undergo the same GRI, we focused on accretion rates higher than our choice of 3$\times$10$^{-3} M_\odot\,\mathrm{yr^{-1}}$ since at this accretion rate, the low WIMP density models ($\rho_\chi = 10^{12}$--$10^{14}\,\mathrm{GeV\,cm^{-3}}$) do not reach the required mass for the effects of GRI to become relevant. Instead, we compare new sets of models with background WIMP density, $\rho_\chi = 10^{13}$ and $10^{15}\,\mathrm{GeV\,cm^{-3}}$ at an accretion rate of 1$\times$10$^{-2} M_\odot\,\mathrm{yr^{-1}}$. 

\begin{figure}[ht]
    \centering
    \includegraphics[width=\columnwidth]{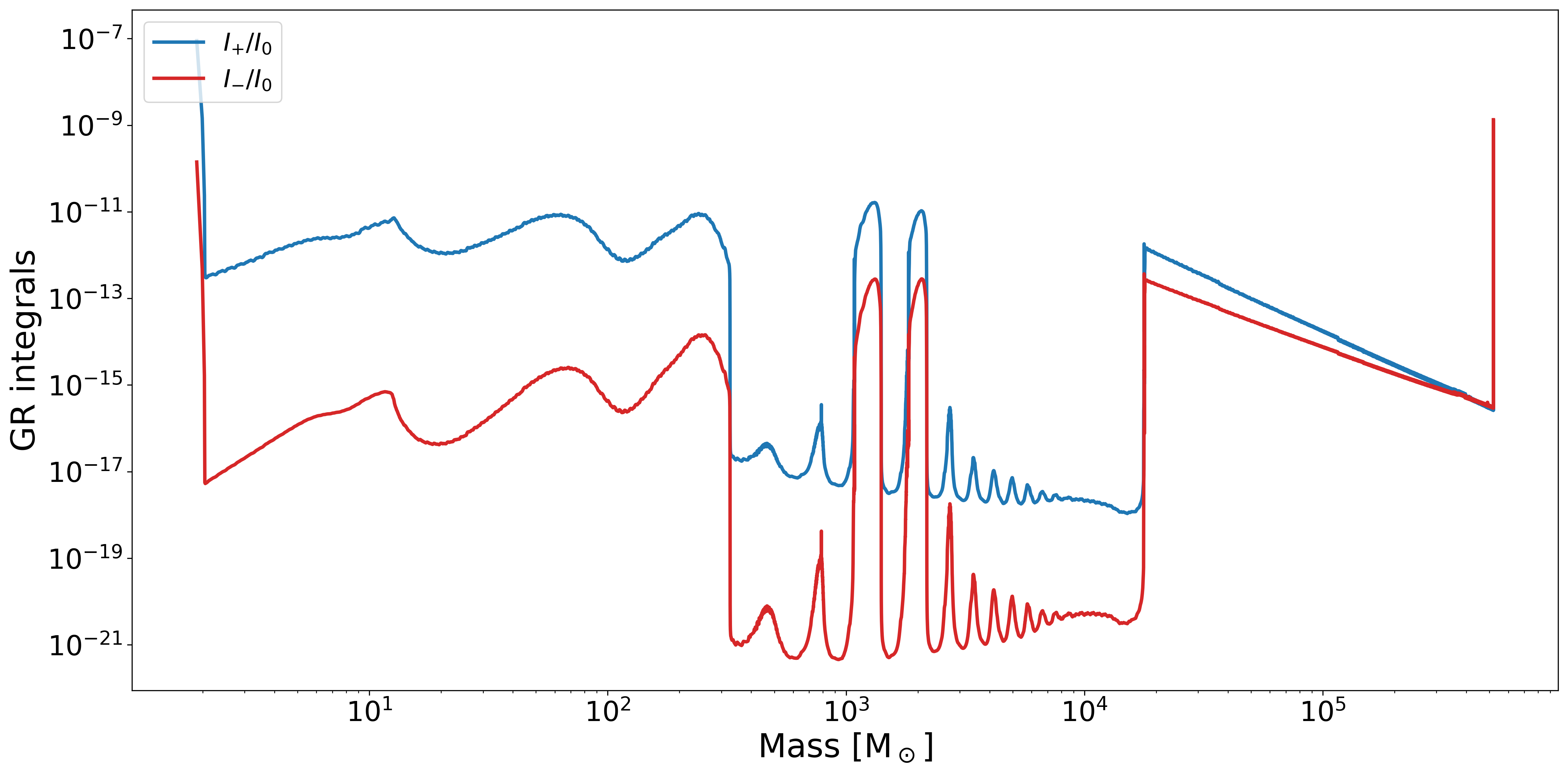}\\[-2pt]
    \includegraphics[width=\columnwidth]{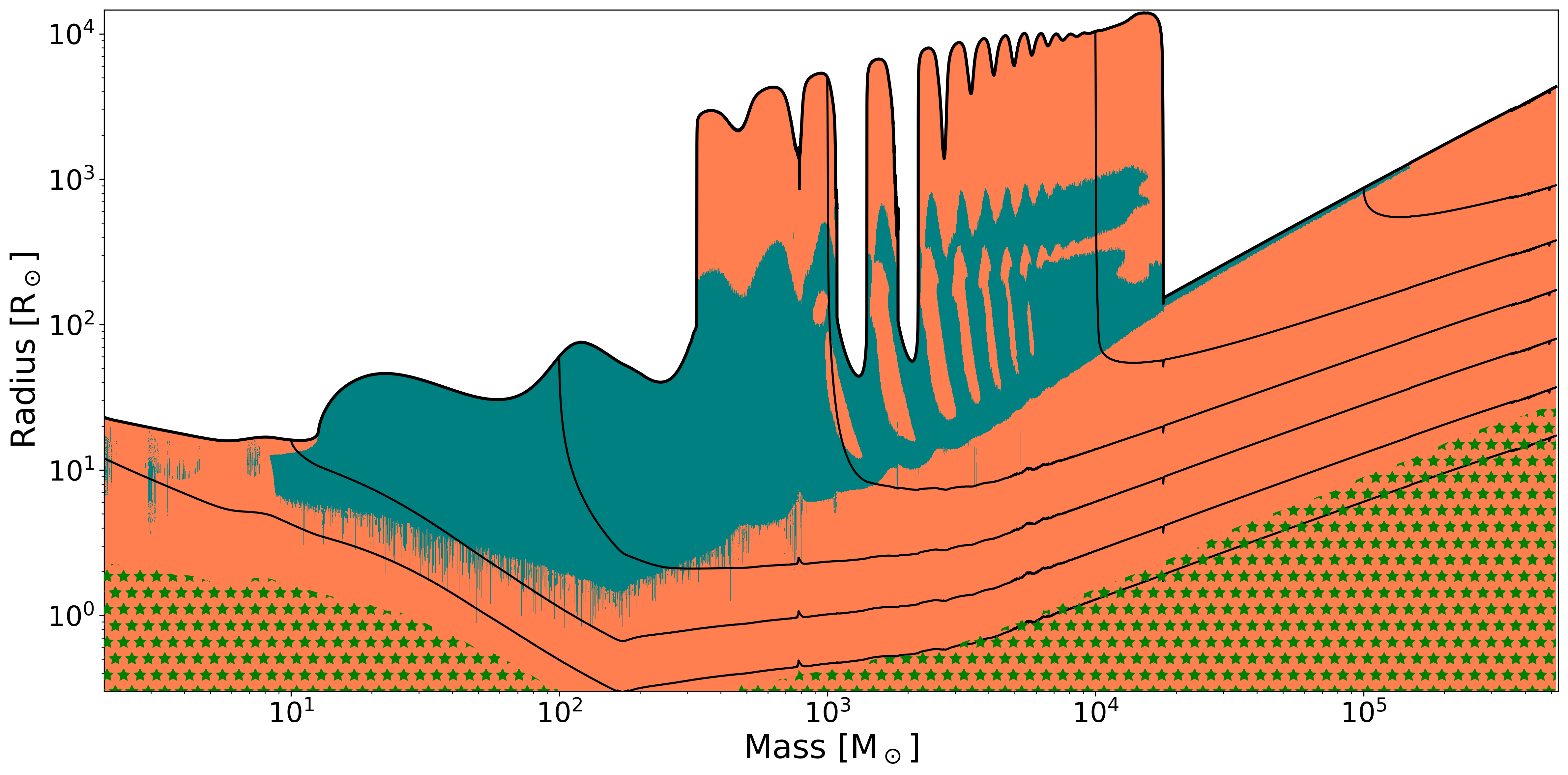}
    \caption{General relativistic integrals (top) and Kippenhahn
             diagram (bottom) for the
             $\rho_{\chi}=10^{13}$\,GeV\,cm$^{-3}$ model.  The GRI is reached when the blue and red curves meet
             at $M_*\simeq4.8\times10^{5}\,M_\odot$.  In the lower panel,
             coral shading denotes convective regions, teal shading
             radiative regions, and green stars WIMP-heating layers.}
    \label{fig:rho13}
\end{figure}

\begin{figure}[ht]
    \centering
    \includegraphics[width=\columnwidth]{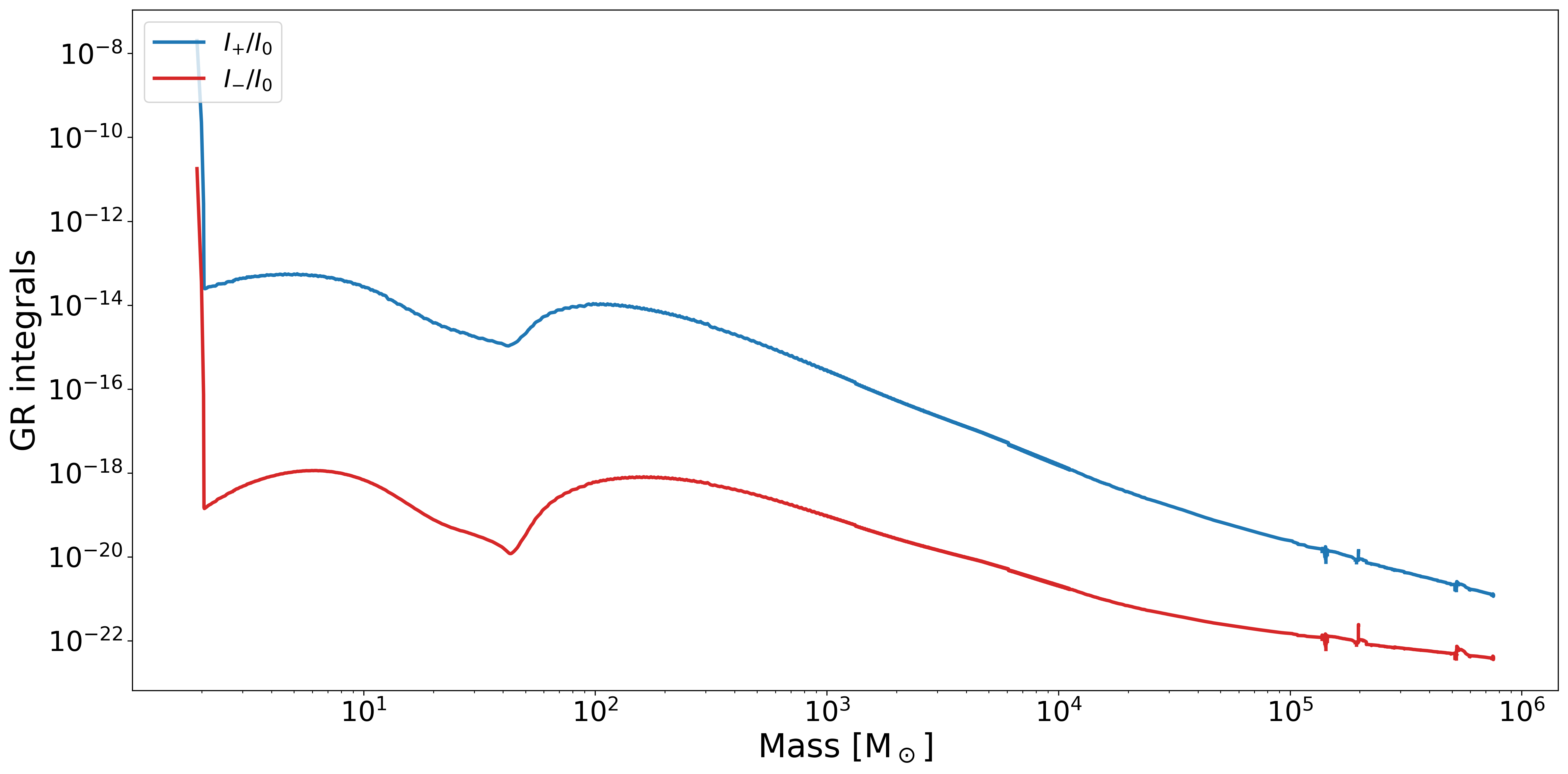}\\[-2pt]
    \includegraphics[width=\columnwidth]{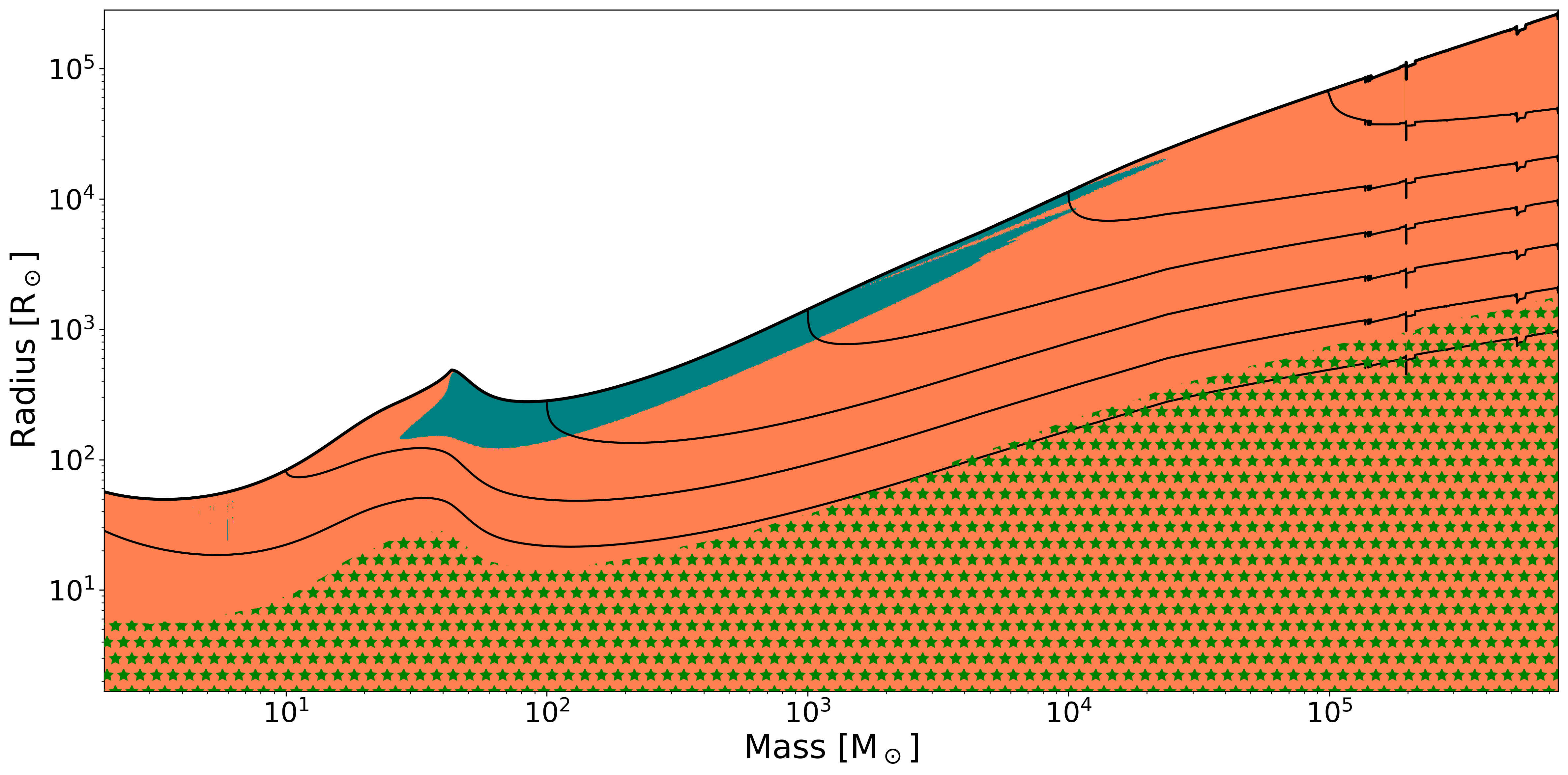}
    \caption{Same as Fig.~\ref{fig:rho13} but for
             $\rho_{\chi}=10^{15}$\,GeV\,cm$^{-3}$.
             The blue and red curves never intersect, so the star
             remains GR-stable up to the final mass
             $M_*\sim10^{6}\,M_\odot$.}
    \label{fig:rho15}
\end{figure}

The top panels of Figs.~\ref{fig:rho13} and \ref{fig:rho15}
show the Chandrasekhar GR-instability integrals
$I_{+}/I_{0}$ and $I_{-}/I_{0}$ for two otherwise identical SMS
models that differ only in their ambient WIMP density. The
quantities $I_{+}$ and $I_{-}$ are defined following
\citet{Chandrasekhar_1964b} and implemented as described by
\citet{Lionel2021}. The star becomes dynamically unstable when
$I_{+}=I_{-}$. The $I_{+}/I_{0}$ and $I_{-}/I_{0}$ curves are
shown in blue and red, respectively.
The corresponding Kippenhahn diagrams are placed underneath each
integral plot.
In these plots the coral shading marks convective regions, the teal shading marks radiative regions, and green stars highlight layers where the
energy budget is dominated by WIMP annihilation.

%----------------------------- 10^13 ---------------------------------
First we discuss the model at background WIMP density of $\rho_{\chi}=10^{13}$\,GeV\,cm$^{-3}$
(Fig.~\ref{fig:rho13}).
During the first $\sim10^{4}\,M_\odot$ of growth the star undergoes
large‐amplitude radius oscillations (seen as the vertical excursions of
the surface isomass line in the lower panel).
These excursions are not due to nuclear flashes---no hydrogen is ignited
in this model---but arise from the “luminosity‐wave’’ mechanism that
operates when the accretion rate
$\dot M_*=10^{-2}\,M_\odot$\,yr$^{-1}$
is a factor ${\sim}4$ below the critical value
$2.5\times10^{-2}\,M_\odot$\,yr$^{-1}$
\citep{Hosokawa_2010,Nandal2024}.
Although the envelope repeatedly inflates and contracts,
$I_{+}/I_{0}$ remains at least an order of magnitude above
$I_{-}/I_{0}$, so the star is still GR stable.

Beyond $M_*\sim10^{4}\,M_\odot$ the oscillations cease,
the envelope settles, and the core contracts steadily while WIMP
annihilation becomes a comparatively minor energy source
(thin green layer).
As the mass increases, the dimensionless compactness
$GM/(Rc^{2})$ grows and the destabilising term $I_{-}$ rises more
quickly than $I_{+}$. The $I_{+}/I_{0}$ and $I_{-}/I_{0}$ curves
shown in Figs.~\ref{fig:rho13} and \ref{fig:rho15} were smoothed
with a log-space boxcar filter to suppress isolated numerical
spikes. When the star attains
$M_*\simeq4.8\times10^{5}\,M_\odot$, the two curves finally
intersect, marking the Chandrasekhar point and the onset of GR
collapse expected for Pop~III supermassive stars with moderate
DM support.
%----------------------------- 10^15 ---------------------------------
Next, we look at the model with background WIMP density, $\rho_{\chi}=10^{15}$\,GeV\,cm$^{-3}$
(Fig.~\ref{fig:rho15}).
Here the DM capture rate is two orders of magnitude larger, and the
green WIMP-heating layer extends over most of the core.
The additional energy input keeps the envelope bloated
($R\gtrsim10^{3}\,R_\odot$), so the star never experiences the
radius oscillations seen in the lower–density case.
Throughout the entire evolution to
$M_*\sim10^{6}\,M_\odot$
the stabilising integral $I_{+}$ exceeds $I_{-}$ by more than an order
of magnitude; the curves never converge, and the Chandrasekhar
criterion is not satisfied.
Consequently the model remains dynamically stable and can in
principle continue accreting beyond the simulated endpoint. To put this result into perspective with our choice of accretion rate (3$\times$10$^{-3} M_\odot\,\mathrm{yr^{-1}}$), we find that models with $\rho_{\chi}=10^{15}$ and $10^{16}$\,GeV\,cm$^{-3}$ do not reach the GRI since their structure is identical to the model depicted in Figure~\ref{fig:rho15}.

%\medskip
These examples illustrate the key role of WIMP annihilation heating in
modulating GR stability.
When the DM reservoir is limited
($\rho_{\chi}=10^{13}$\,GeV\,cm$^{-3}$)
the core eventually contracts enough for relativistic corrections to
dominate, triggering collapse at
$M_*\sim5\times10^{5}\,M_\odot$.
At $\rho_{\chi}=10^{15}$\,GeV\,cm$^{-3}$ the extra heating keeps the
star diffuse, lowers its central density, and prevents the integrals
from meeting—allowing growth to at least
$10^{6}\,M_\odot$ without encountering the GRI. 

%------------------------------------------------------------------

\section{Discussion}\label{Sec:Discussion}

\subsection{Impact of DM depletion on stellar structure}\label{Sec:DMdep}

\begin{figure}[ht]
    \centering
    \includegraphics[width=\columnwidth]{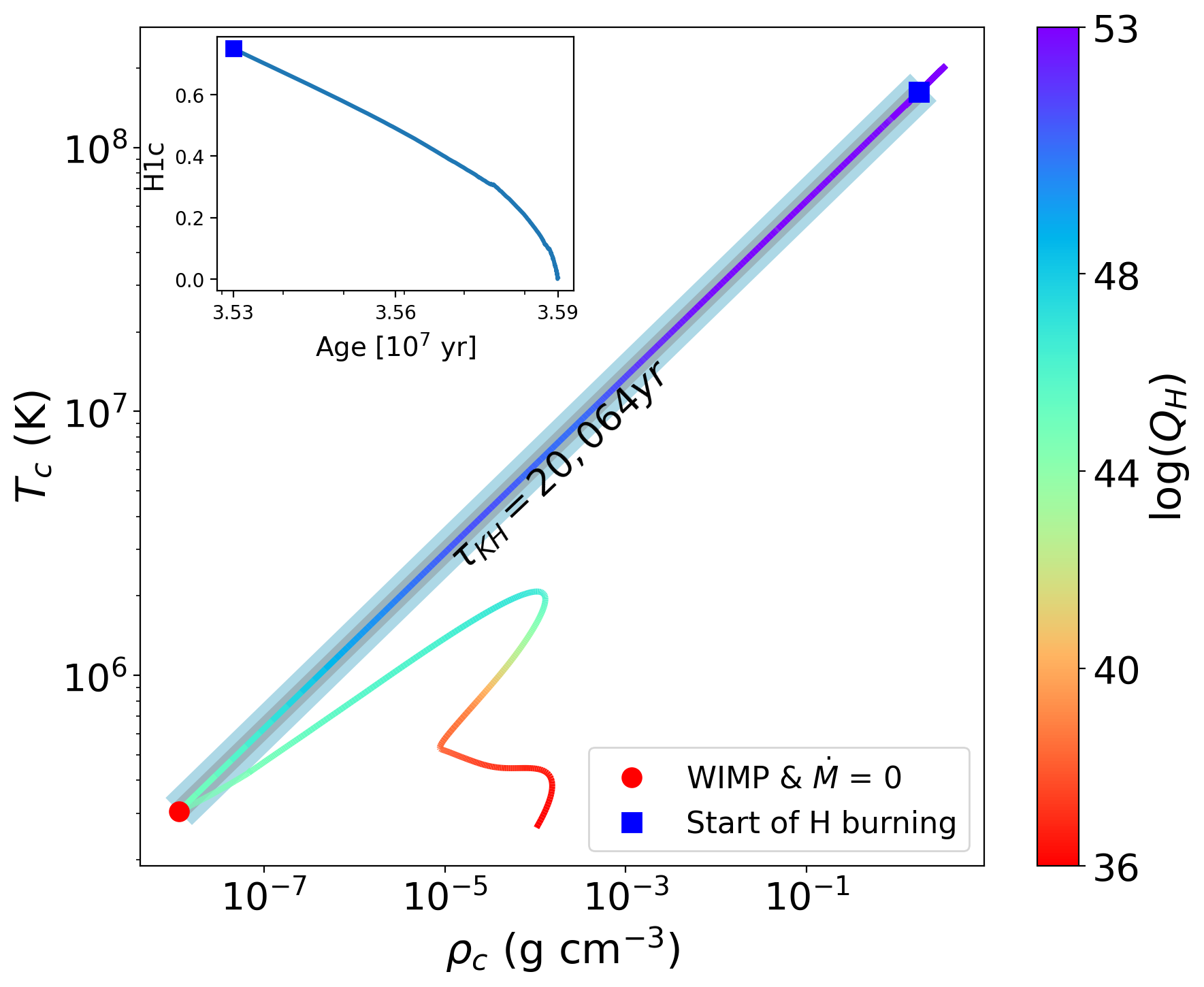}\\
    \caption{Central temperature \(T_{c}\) versus central density \(\rho_{c}\) for a \(10^{5}\,M_\odot\) star accreting at \(10^{-2}\,M_\odot\,\mathrm{yr}^{-1}\) in a WIMP background of \(10^{15}\,\mathrm{GeV\,cm^{-3}}\). Once a mass of 10$^{5}\:M_\odot$ is reached, accretion and DM capture is turned off. The track is coloured by \(\log (Q_H)\).  The red circle indicates the point when accretion and DM capture are turned off; the blue square marks the beginning of core hydrogen burning.  A 20,000 yr light-blue segment and grey connector show the Kelvin–Helmholtz timescale \(\tau_{\rm KH}\).  - Upper-left inset: Post-drop Xc versus age, \(X_c\) vs.\ age in units of \(10^7\) yr.}

    \label{fig:Tcrhoc_wimpoff}
\end{figure}

Here we explore the effects of terminating the DM capture in our fiducial model ($\rho_{\chi}=10^{15}$\,GeV\,cm$^{-3}$). Once the model reaches a mass of about $10^{5} M_\odot$, we assume the model has depleted all of the available baryonic gas and DM reservoir, and we manually terminate accretion and DM capture. This change to the model is initiated at a luminosity, log~($L/L_\odot$) = 9.8 and an effective temperature, log~($T_{\rm eff}$) = 3.75, and at an age of 35.3~Myr (see model (e) in the left panel of Figure~\ref{fig:1_HR_and_tcrhoc}). The stellar structure at this stage is entirely dependent on the DM reservoir, which has kept the central temperature at 3$\times10^{6}\:$K and the central density at 10$^{-8}\:$g~cm$^{-3}$. This position is depicted by the red dot in Figure~\ref{fig:Tcrhoc_wimpoff}. As the capture rate of WIMPS at this stage is zero, the total WIMP reservoir in the star amounts to 0.95~$M_\odot$.

We find the contraction timescale of this model from the termination of WIMP capture (and accretion) until the onset of core hydrogen burning to be around 20,000 years. However, if we were to analytically estimate the contraction timescale ($\tau_{KH}$ of this model, either by simple estimation or by including the effects of Eddington limit, we would obtain  
\begin{equation}
\tau_{\rm KH}^{\rm simp}
= \frac{G\,M^2}{2\,R\,L}
\approx 2.06\times10^{3}\ \mathrm{yr},
\qquad
\tau_{\rm KH}^{\rm Edd}
= \frac{3\,\kappa_{\rm es}\,M}{20\pi\,c\,R}
\approx 2.18\times10^{3}\ \mathrm{yr}.
\end{equation}
Here, we assumed \(M=1.02374\times10^5\,M_\odot\), \(R=2.30\times10^4\,R_\odot\), and electron‐scattering opacity \(\kappa_{\rm es}=0.34\)\,cm\(^2\) g\(^{-1}\). This results in  
\(\tau_{\rm KH}\approx2.18\times10^3\)\,yr which in an order of magnitude below the actual contraction timescale of the model. The discrepancy is due to the effects of DM heating coming from an ever-depleting reservoir of WIMPS in the stellar structure. This DM heating allows the model to stay at its initial position of log~($L/L_\odot$) = 9.8 and an effective temperature, log~($T_{\rm eff}/{\rm K}$) = 3.75 (see model (e) in Figure~\ref{fig:1_HR_and_tcrhoc}) for another 18,000 years before the contraction towards the ZAMS begins. Core hydrogen burning begins at an age of 35.3 Myr, at log~($T_{\rm eff}/{\rm K}$) = 4.5, as shown in the top left panel window of Figure~\ref{fig:rho15}. The evolution continues until the central mass fraction of hydrogen reaches 0.2, at which point the model experiences the GR-instability. The final age of the model at this stage is 35.8 Myr, implying once accretion and DM capture is terminated, the model only survives for another 0.5 Myr.

Another consequence of DM depletion and the subsequent core hydrogen burning is the sharp increase in the number of ionising photons ($Q_{\rm H}$). As the central mass fraction of hydrogen reaches 0.36, the log~($Q_{\rm H}$) rises to about 53, as shown by the colourbar of Figure~\ref{fig:Tcrhoc_wimpoff}. This is also evident in track (ii) in Figure~\ref{fig:QH_accretion} once it reaches a mass of 10$^5\:M_\odot$. The model contracts and consequently reaches a much higher log($T_{\rm eff}/{\rm K}) > 4.8$, which leads to an increase in the number of ionising photons. This high photon flux of $>10^{53}\:{\rm s}^{-1}$ is maintained by the stellar atmosphere for about 0.16~Myr, until the end of the star's evolution. Additionally, the $Q_{H}$ value of this model during core hydrogen burning is always higher than 10$^{51}\:{\rm s}^{-1}$ for around 0.5 Myr. This implies that once the DM reservoir of a Pop~III.1 star depletes, their subsequent contraction towards the ZAMS makes them a strong source of ionising photons for the rest of their evolutionary stages. 

\subsection{Accretion rate, dark matter build‑up, and the
           ionising flash}\label{Sec:AccRate_QH}

\begin{figure}
  \centering
  \includegraphics[width=\columnwidth]{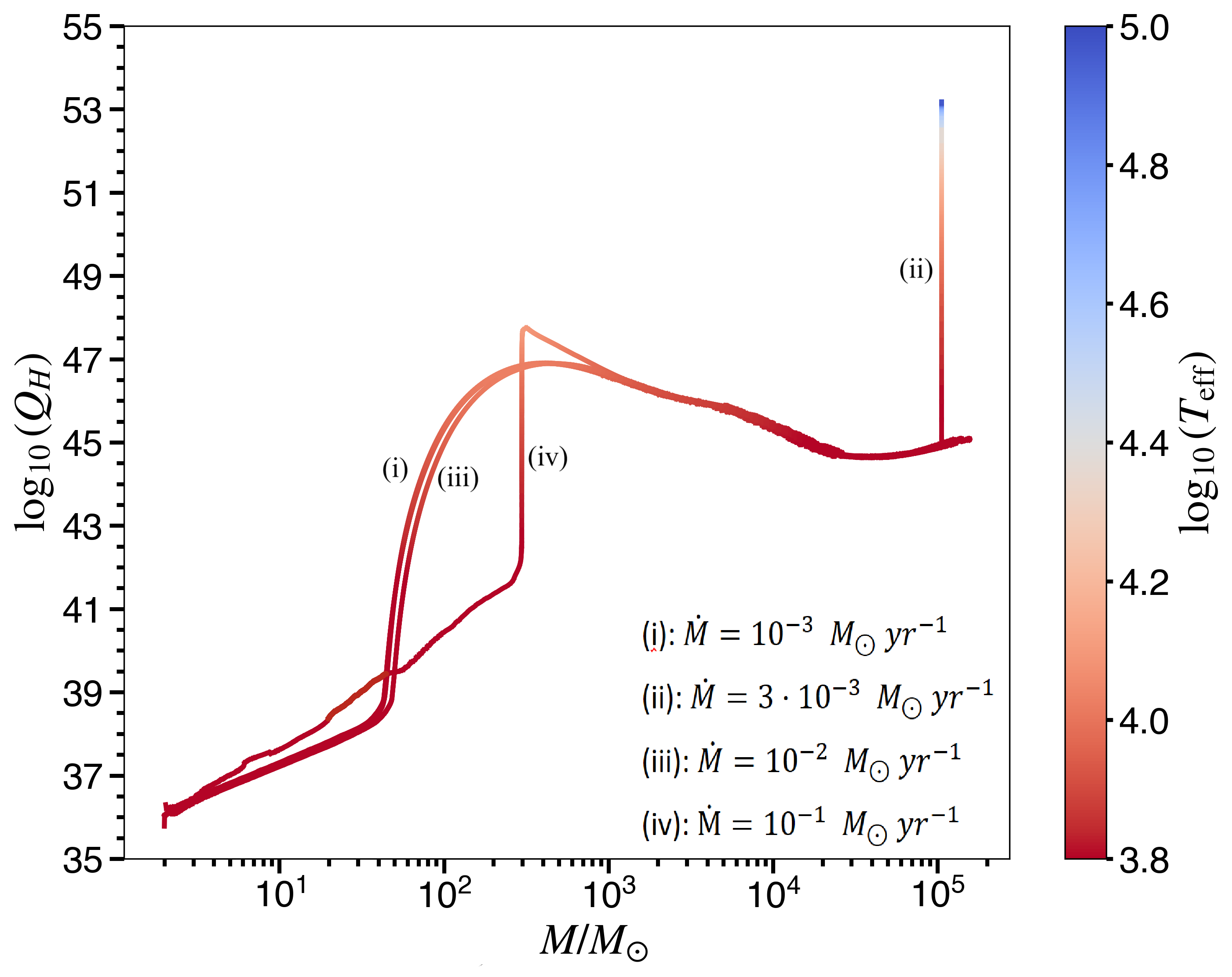}
  \caption{Hydrogen‑ionising photon rate for supermassive stars accreting in a
           $\rho_\chi = 10^{15}\,\mathrm{GeV\,cm^{-3}}$ environment.  Accretion
           rates are colour‑coded by $\log (T_\mathrm{eff}/{\rm K})$; labels (i), (ii), (iii), and (iv) correspond to models with accretion rates of 10$^{-3}$, $3\times10^{-3}$, 10$^{-2}$, and 10$^{-1}\:M_\odot\:{\rm yr}^{-1}$, respectively.}
  \label{fig:QH_accretion}
\end{figure}

Figure~\ref{fig:QH_accretion} shows the evolution of the hydrogen‑ionising
photon rate, $Q_H$, for four supermassive protostars that grow in the same
dark‑matter environment
($\rho_\chi = 10^{15}\,\mathrm{GeV\,cm^{-3}}$) but accrete at
$\dot M = 10^{-1}$, $10^{-2}$, 3$\times$10$^{-3}$ and
$10^{-3}\,M_\odot\,\mathrm{yr^{-1}}$. At the start of the computation all three models are cool ($\log (T_\mathrm{eff}/{\rm K})\simeq3.65$) and inflated ($R\sim10^{3}$–$10^{4}\,R_\odot$), so their initial ionising output is negligible ($\log (Q_H/{\rm s}^{-1})\lesssim36$).

The subsequent behaviour depends on how quickly baryonic mass is added compared with how quickly the WIMP reservoir can grow.  In the run with $\dot{M}_* = 0.1\,M_\odot\,\mathrm{yr^{-1}}$ the star reaches $M_*\simeq2.5\times10^{2}\,M_\odot$ in only $\simeq2.5\times10^{3}\,$yr.  At this point the photon luminosity has risen to $\log L/L_\odot\sim 7$, yet the central WIMP density has not increased enough to provide full support. The envelope therefore contracts on a Kelvin–Helmholtz timescale and raises the surface temperature to $\log (T_\mathrm{eff} /{\rm K})\simeq4.3$. Since $Q_H$ scales exponentially with $T_\mathrm{eff}$ in the Wien tail, this brief contraction produces a narrow ionising flash peaking at $\log Q_H\approx47.8$.  Within a few thousand years the capture rate catches up, dark‑matter heating once again balances radiative losses, and the star re‑expands to a cooler state.
 
The tracks with $\dot{M}_* = 10^{-2}$, 3$\times$10$^{-3}$, and $10^{-3}\,M_\odot\,\mathrm{yr^{-1}}$ grow more slowly
($\gtrsim2.5\times10^{4}$ and $\gtrsim2.5\times10^{5}$\,yr,
respectively).  During this longer pre‑main‑sequence phase the central WIMP density builds up smoothly, compensating the growing photon luminosity before contraction can set in.  As a result the envelopes remain extended, $\log (T_\mathrm{eff}/{\rm K})$ never exceeds $\simeq4.05$, and $Q_H$ rises monotonically to a common plateau $\log (Q_H/{\rm s}^{-1})\simeq46.2\pm0.2$ between $M_*\sim10^{3}$ and $10^{5}\,M_\odot$. Once the stars approach $M_*\gtrsim10^{3.5}\,M_\odot$, heating by WIMP annihilation dominates in all four cases.  The envelopes undergo a slow secular expansion, $T_\mathrm{eff}$ decreases gradually, and $Q_H$ decreases.

The comparison highlights a clear dichotomy.  If the accretion rate is moderate or slow, the dark‑matter reservoir has time to grow and the star remains in a bloated, weakly ionising state throughout its life. Only when mass is supplied at near‑maximal rates does the protostar contract briefly, forming a compact, UV‑bright configuration that generates an ionising flash
of $\log Q_H\sim48$. 

\subsection{Comparison with previous studies}\label{Sec:CompPrev}

A growing body of work has examined how WIMP-baryon interactions reshape primordial star formation. Broadly, the literature splits into two camps: (i) continuous–capture models, which assume that adiabatic contraction or persistent scattering keeps replenishing the stellar WIMP reservoir, and (ii) finite–reservoir models, in which capture cannot keep pace with stellar growth, so WIMP support wanes over time. Our grid includes both regimes: the default tracks keep capturing WIMPs as the star grows to ${\sim}10^{5}\,M_\odot$, whereas a dedicated ``WIMP–off’’ run switches capture off at that mass to isolate the post‑DM phase (Sect.~\ref{Sec:DMdep}).

Our models of DM-powered protostars exhibit a general‐relativistic (GR) instability at final masses of order $10^5$–$10^6\,M_\odot$, in line with the upper envelope obtained by \citet{Haemmerle2024}. This author showed that as long as DM heating maintains $T_c\lesssim10^7$\,K, the Chandrasekhar integral $I_+/I_0$ stays above $I_-/I_0$ until $M_*\gtrsim10^6\,M_\odot$.  By contrast, non–rotating polytropes without DM support collapse already at $(2$–$4)\times10^5\,M_\odot$ \citep{Baumgarte1999}.  Our $\rho_\chi=10^{13}\:{\rm GeV\:cm}^{-3}$ model reaches the GR point at $M_*\simeq4.8\times10^5\,M_\odot$, whereas the $\rho_\chi=10^{15}\:{\rm GeV\:cm}^{-3}$ track grows to $8\times10^5\,M_\odot$ without instability, matching the trend that stronger WIMP heating pushes the critical mass upward.

A principal difference of our work compared to that of \citet{RindlerDaller2015} lies in the capture prescription.  Under strong adiabatic contraction they find that WIMP annihilation remains the dominant energy source throughout the entire growth to $10^5\,M_\odot$, keeping their stars relatively cool 
%($T_{\rm eff}\sim5\times10^3$\,K) 
and preventing GR collapse.  When we impose the same halo profile our stars behave similarly, but in the baseline runs we adopt the weaker, single‑scatter capture of \citet{Gould1987}. 
%the resulting 
%finite reservoir 
For the cases where we assume WIMP capture is exhausted after $\sim10^5\:M_\odot$, we see contraction and eventual GRI during the core hydrogen burning stage. Hence the contrasting final masses are rooted not in numerics but in the choice of halo evolution and cross‑sections.

The ionising photon output $Q_H$ in our models rises very steeply once the star contracts and heats up.  After a brief pre‑main‑sequence phase where $\log Q_H\lesssim36$, all tracks with $\dot{M}_*\le10^{-2}\,M_\odot\,\mathrm{yr^{-1}}$ converge to a plateau $\log Q_H\simeq46.2\pm0.2$ over $10^{3}\lesssim M_*/M_\odot\lesssim10^5$.  \citet{Ilie2021}, who adopted the larger spin‑independent cross‑sections then allowed by XENON1T, found $Q_H$ to remain $<10^{40}\,\mathrm{s^{-1}}$ until $M_*\gtrsim10^5\,M_\odot$ because their envelopes never contracted.  The difference again traces back to whether DM capture can keep the envelope inflated.

A related observable is the radius–mass relation.  \citet{Wu2022} examined self‑interacting DM and reported that SIDM heating increases $R$ by at most a factor of two relative to collisionless cases.  We reproduce a similar modest effect: once nuclear burning ignites, $R(M_*)$ in our $\rho_\chi=10^{15}\:{\rm GeV\:cm}^{-3}$ track lies only $\sim1.5$\,dex above the ZAMS line, far smaller than the $R\sim100$\,AU predicted for continuously captured WIMP stars \citep{RindlerDaller2015}.  This confirms that envelope inflation scales with the integrated DM energy release rather than the specific particle physics channel.

Overall, the comparison shows that GRI mass, $Q_H(M)$, and $R(M)$ hinge on three modelling choices: (i) whether the halo undergoes strong adiabatic contraction, (ii) the adopted WIMP scattering cross‑sections, and (iii) the treatment of multi‑scatter capture.  Within current experimental bounds, our fixed background density models reproduce the high‑temperature, high‑$Q_H$ behaviour expected of massive Pop~III stars while still delaying GR collapse to $\gtrsim5\times10^5\,M_\odot$—a regime intermediate between Hayashi-limited tracks and pure‑fusion tracks.

\subsection{Implications for critical WIMP density for SMS formation}

One of the main results of our study is the identification of a critical ambient WIMP density required for a Pop III.1 protostar to grow to supermassive scales. For a fiducial gas accretion rate of $\dot{M}_{*} = 3 \times 10^{-3} \, M_{\odot} \, \text{yr}^{-1}$, this threshold density is $\rho_{\chi} \gtrsim 5 \times 10^{14} \, \text{GeV} \, \text{cm}^{-3}$. Below this density, ionising feedback truncates accretion, limiting the final stellar mass to a few hundred solar masses, far short of the $\gtrsim 10^5 \, M_{\odot}$ required for heavy black hole seeds.

It remains uncertain whether such DM densities can be achieved in Pop III.1 minihalos. \citet{abn02} and \citet{2009ApJ...692..574N} reported evidence of adiabatic contraction in numerical simulations of Pop III.1 star formation. In particular, in the inner regions ($\lesssim0.5\:$pc), where the density of baryons begins to dominate over that of DM, the radial profile of DM density was observed to steepen to follow a power law index close to that of the gas, i.e., $\rho_\chi \propto r^{-\alpha_\chi}$ with $\alpha_\chi\simeq 2$.

\citet{2012ApJ...761..154S} simulated Pop III.1 star formation assuming an analytic DM density profile that responds to changes in peak gas density, with central values at radii of $\sim1\:$au being $\sim10^{13}\:{\rm GeV\:cm}^{-3}$. They concluded that the heating effects of DM annihilation help stabilise the protostellar accretion disk against fragmentation. \citet{2014MNRAS.441..822S} found similar results, but also noted that in their model in which the DM responded to gravitational interactions with the gas that its density could be reduced to values of $\sim 10^{11}\:{\rm GeV\:cm}^{-3}$ on these scales.

\citet{HiranoYoshida2025} simulated the concentration of DM self-consistently in collapse of Pop III.1 minihalos. In their fiducial halo they found relatively low DM densities of $\sim 10^{5}\:{\rm GeV\:cm}^{-3}$ on scales of $\sim 0.01\:$pc (2,000~au) (with this inner scale set by application of a gravitational softening on scales of 0.022~pc), with a shallow power lar profile that would predict densities of $\sim 10^{7}\:{\rm GeV\:cm}^{-3}$ at $\sim1\:$au. However, this halo had a relatively cool central temperature of $\sim 100\:$K and developed a very large central disk with spiral arms, leading to rotational support out to scales of $\sim 10\:$pc. Such a configuration is not expected to lead to efficient adiabatic contraction of DM.
We speculate that in warmer halos, where central temperatures reach $\sim 1000$~K, the gas structure is more spherical and less perturbed by interaction with spiral arms and stronger degrees of adiabatic contraction may occur.

%These limitations mean their results do not necessarily preclude the high DM densities considered in our models.

\section{Conclusions}\label{Sec:Conclusions}

We have introduced the Gould DM capture formalism into the accreting supermassive star branch of the \textsc{GENEC} stellar evolution code \citep{Eggenberger2008, Nandal2024} and performed stellar evolutionary calculations of DM-powered protostars. By incorporating detailed DM capture and annihilation physics, we systematically explored the impact of varying background DM densities ($\rho_\chi = 10^{12}-10^{16}\,\mathrm{GeV\,cm^{-3}}$) and accretion rates ($\dot{M}_* = 10^{-3}-10^{-1}\,M_\odot\,\mathrm{yr^{-1}}$) on stellar structure, stability, luminosity budgets, and ionising photon outputs. We found that the inclusion of DM capture significantly modifies the evolutionary trajectories and final states of DM-powered stars, highlighting distinct observational signatures directly linked to their DM environments. Our key findings can be summarised as follows:

\begin{enumerate}
    \item A critical dark matter density for supermassive growth. The principal finding is the identification of a critical ambient WIMP density required for a DM-powered protostar to grow to supermassive scales. For a fiducial gas accretion rate of $\dot{M}_{*} = 3 \times 10^{-3} \, M_{\odot} \, \text{yr}^{-1}$, this threshold density is $\rho_{\chi} \gtrsim 5 \times 10^{14} \, \text{GeV} \, \text{cm}^{-3}$. Below this density, powerful ionising feedback halts accretion, limiting the final stellar mass to a few hundred solar masses, far short of the $\gtrsim 10^5 \, M_{\odot}$ required for heavy black hole seeds.

    \item Stellar inflation and feedback suppression. In high-density dark matter environments ($\rho_{\chi} > 5 \times 10^{14} \, \text{GeV} \, \text{cm}^{-3}$), the WIMP annihilation luminosity ($L_{\chi}$) becomes the dominant energy source. This powerful internal heating inflates the stellar envelope to radii of $\sim 10$ AU and lowers the effective surface temperature to a cool $\sim 10^4 \, \text{K}$. Consequently, the ionising photon output ($Q_H$) is quenched by orders of magnitude during the main growth phase, thereby permitting uninterrupted accretion to masses exceeding $10^5 \, M_{\odot}$.

    \item Dark matter as a regulator of general relativistic stability. The ambient DM density is a crucial factor in determining the star's susceptibility to the general relativistic radial instability (GRRI). In environments with moderate DM densities ($\rho_{\chi} \leq 10^{13} \, \text{GeV} \, \text{cm}^{-3}$), stars become unstable and collapse at masses of $\sim(4-5) \times 10^5 \, M_{\odot}$. In contrast, for the highest densities considered ($\rho_{\chi} \geq 10^{15} \, \text{GeV} \, \text{cm}^{-3}$), enhanced DM heating keeps the star's structure sufficiently diffuse to remain stable against GRRI to masses beyond $10^6 \, M_{\odot}$. This mechanism directly links the properties of the host DM halo to the initial mass of the resulting black hole seed.

    \item The interplay of accretion and dark matter capture. The stellar evolution is sensitive to the balance between the gas accretion rate ($\dot{M}_{*}$) and the DM capture rate. High gas accretion rates (e.g. $\dot{M}_{*} = 0.1 \, M_{\odot} \, \text{yr}^{-1}$) can temporarily outpace the buildup of the internal DM reservoir, causing brief periods of stellar contraction and enhanced ionising radiation before the star re-inflates as the DM heating re-establishes dominance. This predicts a `stuttering' growth phase with fluctuating observational properties.

    \item The post-growth luminous phase. A phase of extremely high ionising luminosity is triggered only after the continuous capture of WIMPs from the halo ceases, allowing the star to contract gravitationally. This process heats the star to the Zero-Age Main Sequence, producing an ionising photon output as high as $Q_H \approx 10^{53} \, \text{s}^{-1}$ that is sustained for approximately 0.5 Myr. This powerful radiation would ionise a large surrounding region, creating a giant HII region that serves as a distinct observational signature.

    \item Distinct observational signatures. The co-evolution of the star and its DM halo produces distinct evolutionary phases. The primary growth phase is characterised by a cool ($\sim 10^4 \, \text{K}$), bloated ($\sim 10$ AU), and extremely luminous ($L \sim 10^9 - 10^{10} \, L_{\odot}$) star with weak ionising output, corresponding to a `red' supermassive star phase. This is followed by a potential long-lived ($\sim 0.5$ Myr), hyper-luminous main-sequence phase ($Q_H \approx 10^{53} \, \text{s}^{-1}$), analogous to a `blue' supermassive star, once the external DM fuel supply is exhausted, offering a testable signature for high-redshift surveys with facilities like the JWST \citep{Surace2018, Surace2019}.
\end{enumerate}

Our work demonstrates that the evolution of DM-powered protostars is deeply intertwined with their DM environment, underscoring the importance of considering DM physics in the formation and evolution of primordial stars. The distinct evolutionary paths and observational signatures, such as prolonged protostellar expansion phases and the delayed onset of GRI, provide the foundation for making predictions that may be testable by future high-redshift observational campaigns, particularly with facilities like JWST.

Future investigations should address open questions by extending the present framework. Promising directions include: (i) incorporating the effects of adiabatic contraction using the Blumenthal formalism \citep{Blumenthal1986,Freese2009} to better represent evolving halo density profiles; (ii) conducting detailed parameter studies covering broader ranges of accretion rates, DM densities, WIMP masses, and scattering cross-sections (spin-dependent and spin-independent; \citealt{LZ2024}); (iii) exploring the influence of rotation on stellar structure, angular momentum transport, and stability criteria \citep{Ekstrom2012}; and (iv) examining the role of internal magnetic fields in modifying angular momentum distribution, convective stability, and ionising photon output \citep{Nandal2024b}. We aim to explore the sensitivity of the results to WIMP properties (Topalakis et al., in prep.) in Paper II and we will carry out a more thorough exploration of the Post-Growth Luminous Phase, including detailed predictions for observational signatures (Sergienko et al., in prep.) in Paper III.

\begin{acknowledgements}
D.N. acknowledges support from a VITA-Origins postdoctoral fellowship. J.C.T. acknowledges support from ERC Advanced Grant MSTAR (788829) and funding from the Virginia Institute for Theoretical Astrophysics (VITA), supported by the College and Graduate School of Arts and Sciences at the University of Virginia. 
%The authors would like to thank....
\end{acknowledgements}

\bibliographystyle{aa}
\bibliography{biblio}

\end{document}